\documentstyle[12pt,preprint,aps,floats]{revtex}

\tighten

\input epsf.tex

\def\NPB#1#2#3{Nucl. Phys. {\bf B#1}, #2 (19#3)}

\def\PLB#1#2#3{Phys. Lett. {\bf B#1}, #2 (19#3)}
\def\PLBold#1#2#3{Phys. Lett. {\bf#1B}, #2 (19#3)}

\def\PRD#1#2#3{Phys. Rev. {\bf D#1}, #2 (19#3)}
\def\PRL#1#2#3{Phys. Rev. Lett. {\bf#1}, #2 (19#3)}

\newcommand{\postscript}[2]{\setlength{\epsfxsize}{#2\hsize}
   \centerline{\epsfbox{#1}}}
\newcommand{\sign}{\:\!\text{sign}\:\!}
\newcommand{\mgaugino}{M_{1/2}}
\newcommand{\mt}{m_t}
\newcommand{\minit}{Q_0}
\newcommand{\mgut}{M_{\text{GUT}}}
\newcommand{\mplanck}{M_{\text{Pl}}}
\newcommand{\tb}{\tan\beta}
\newcommand{\bold}[1]{\mbox{\boldmath $#1$}}
\newcommand{\ifb}{ \text{ fb}^{-1}}
\newcommand{\gev}{\text{ GeV}}
\newcommand{\tev}{\text{ TeV}}

\begin{document}

\draft

\renewcommand{\thefootnote}{\fnsymbol{footnote}}
\setcounter{footnote}{0}

\preprint{
\noindent
\hfill
\begin{minipage}[t]{3in}
\begin{flushright}
IASSNS--HEP--99--81\\
FERMILAB--PUB--99/248--T\\
hep-ph/9909334\\
September 1999\\
\end{flushright}
\end{minipage}
}

\title{
\vskip 0.5in
Focus Points and Naturalness in Supersymmetry}

\author{
Jonathan L.~Feng$^a$, Konstantin T.~Matchev$^b$, and Takeo
Moroi$^a$
\vskip 0.2in
}

\address{
  ${}^{a}$
  School of Natural Sciences,
  Institute for Advanced Study\\
  Princeton, NJ 08540, U.S.A. 
\vskip 0.1in
  ${}^{b}$
  Theoretical Physics Department,
  Fermi National Accelerator Laboratory\\
  Batavia, IL 60510, U.S.A.}

\maketitle

\begin{abstract}

We analyze focus points in supersymmetric theories, where a
parameter's renormalization group trajectories meet for a family of
ultraviolet boundary conditions.  We show that in a class of models
including minimal supergravity, the up-type Higgs mass has a focus
point at the weak scale, where its value is highly insensitive to the
universal scalar mass.  As a result, scalar masses as large as 2 to 3
TeV are consistent with naturalness, and {\em all} squarks, sleptons
and heavy Higgs scalars may be beyond the discovery reaches of the
Large Hadron Collider and proposed linear colliders.  Gaugino and
Higgsino masses are, however, still constrained to be near the weak
scale.  The focus point behavior is remarkably robust, holding for
both moderate and large $\tb$, any weak scale gaugino masses and $A$
parameters, variations in the top quark mass within experimental
bounds, and for large variations in the boundary condition scale.

\end{abstract}



\newpage

\renewcommand{\thefootnote}{\arabic{footnote}}
\setcounter{footnote}{0}

\section{Introduction}

An understanding of electroweak symmetry breaking is currently one of
the most important objectives in high energy particle physics.
Renormalizability requires that electroweak symmetry be spontaneously
broken; in the minimal standard model, this is realized by the
condensation of the elementary Higgs field.  In such a theory,
however, the squared Higgs mass receives quadratically divergent
radiative corrections.  The Higgs mass, and with it the weak scale, is
therefore expected to be of order the cut-off scale, which is
typically identified with the grand unified theory (GUT) or Planck
scale.  The fact that the weak scale is much smaller than the cut-off
scale requires a large fine-tuning and is therefore considered
unnatural in the minimal standard model~\cite{SM}.

Supersymmetry removes quadratic divergences and therefore provides a
framework for naturally explaining the stability of the weak scale
with respect to radiative corrections~\cite{SUSY}.  However, the
requirement of naturalness constrains supersymmetric models, as, in
these models, the weak scale is generated when electroweak symmetry
breaking is induced by (negative) squared mass parameters for the
Higgs scalars.  The weak scale is therefore related to these
supersymmetry breaking parameters.  It is hoped that an understanding
of the mechanism of supersymmetry breaking will shed light on the
origin of the weak scale.  Even without this knowledge, though, it is
clear that naturalness in supersymmetric theories requires that the
supersymmetry breaking parameters in the Higgs potential be not too
far above the weak scale.

As naturalness implies {\em upper} bounds on supersymmetry breaking
parameters and superpartner masses, its implications are obviously of
great importance for supersymmetry searches.  These implications
depend on the assumed structure of supersymmetry breaking.  With
respect to the scalar masses, broadly speaking, three possibilities
exist.  In the first, all supersymmetry breaking scalar mass
parameters are roughly of the same order; for example, they may be of
the same order when generated at some high scale and then remain of
the same order when evolved through renormalization group (RG)
equations to the weak scale.  Naturalness then demands that scalar
Higgs, squark, and slepton masses all be near the weak scale.  Such
light particles are within the discovery reach of the Large Hadron
Collider (LHC) and future lepton colliders.

Another possibility is that a hierarchy exists between the various
scalar masses.  This hierarchy may be present at the scale at which
supersymmetry breaking parameters are generated~\cite{moreminimal} or
may be generated dynamically through RG evolution~\cite{fkp}.  In
either case, one finds that naturalness bounds on the first and second
generation squarks and sleptons are much weaker than those for the
third generation~\cite{nonuniv}.  First and second generation
sfermions may then be much heavier than 1 TeV and far beyond the reach
of near future colliders.  However, top and bottom squarks, for
example, are still constrained to have masses of order the weak scale,
and should be discovered by the LHC.

A third possibility, however, is that the RG trajectories of the Higgs
mass parameters may meet at a ``focus point''~\cite{anom,paper1},
where their values are independent of their ultraviolet boundary
values.\footnote{Focus points are not be confused with the well-known
phenomena of fixed and quasi-fixed points~\cite{fp}.  As we will see,
when RG trajectories have a focus point behavior, they do not
asymptotically approach a limit curve, but rather meet and then
disperse.}  If this focus point is near the weak scale, the Higgs
potential at the weak scale may be insensitive to the ultraviolet
values of certain supersymmetry breaking parameters, including the
scalar masses.  In this case, naturalness, while constraining
(unphysical) Higgs mass parameters, may lead to very weak upper bounds
on the squark, slepton, and heavy Higgs boson masses, and these
scalars may {\em all} be beyond the reach of near future colliders.

The last possibility is the subject of this study.  We will show that
it is realized in a class of models that includes minimal
supergravity.  Minimal supergravity is at present probably the single
most widely-studied framework for evaluating the potential of new
experiments to probe physics beyond the standard model.  We therefore
concentrate on this model and explore the implications of focus points
in minimal supergravity for naturalness and the superpartner spectrum.

The organization of this paper is as follows.  In
Sec.~\ref{sec:focus}, we analyze the focus point behavior of the RG
evolution of supersymmetry breaking parameters.  In
Sec.~\ref{sec:natural}, we discuss the implications of the up-type
Higgs focus point for naturalness in minimal supergravity.  In
particular, we will see that multi-TeV scalar masses are consistent
with naturalness.  The implications of these results for superpartner
searches are considered in Sec.~\ref{sec:implications}.  Finally, we
conclude in Sec.~\ref{sec:summary} with a summary of our results and
some philosophical comments concerning the concept of naturalness.

\section{Focus Points}
\label{sec:focus}

In this section, we explore the phenomenon of focus points in the RG
evolution of supersymmetry breaking parameters.  We will see that, in
certain circumstances, the supersymmetry breaking up-type Higgs mass
has such a focus point at the weak scale, where it becomes insensitive
to its boundary value at the high scale (for example, the GUT scale).
Since the supersymmetry breaking masses for the Higgses are related to
the weak scale, this fact has implications for naturalness, as we will
discuss in Sec.~\ref{sec:natural}.

We start by considering the RG behavior of supersymmetry breaking
scalar masses.  Denoting the mass of the $i$-th scalar field by $m_i$,
the one-loop RG evolution of scalar masses is given
schematically\footnote{In Eqs.~(\ref{dm^2/dt})--(\ref{ARGE}) and
(\ref{ddm^2/dt}), we neglect positive ${\cal O}(1)$ coefficients for
each term.} by the inhomogeneous equations

\begin{eqnarray}
\frac{dm_i^2}{d \ln Q} \sim \frac{1}{16\pi^2} \biggl[ -g_a^2 M_a^2
+ \sum_j y_j^2 m_j^2 + \sum_j y_j^2 A_j^2 \biggr] \ ,
\label{dm^2/dt}
\end{eqnarray}
where $g_a$ and $y_j$ are gauge and Yukawa coupling constants,
respectively, $Q$ is the renormalization scale, and the summation is
over all chiral superfields coupled to the $i$-th chiral superfield
through Yukawa interactions.  The gaugino masses $M_a$ and
supersymmetry breaking trilinear scalar couplings $A_j$ have RG
equations
\begin{eqnarray}
	\frac{dM_a}{d \ln Q} &\sim& \frac{1}{16\pi^2} g_a^2 M_a \ ,
\\
	\frac{dA_i}{d \ln Q} &\sim& \frac{1}{16\pi^2}
	\biggl[ g_a^2 M_a + \sum_j y_j^2 A_j \biggr] \ .
\label{ARGE}
\end{eqnarray}

As one can see, the evolution of the $m_i^2$ parameters depends on the
gaugino masses and $A$ parameters, as well as on the scalar masses
themselves.  On the other hand, the gaugino masses and $A$ parameters
evolve independently of the scalar masses.\footnote{Note that this
implies that if a hierarchy $M_a, A_j \ll m_i$ is generated at some
high scale, for example, by an approximate $R$-symmetry, it will not
be destabilized by RG evolution.}  This structure implies that, if
$m_i^2|_{\rm p}$ is a particular solution to Eq.~(\ref{dm^2/dt}) with
fixed values of the gaugino masses and $A$ parameters, then for
arbitrary constant $\xi$,
\begin{eqnarray}
	m_i^2(Q) = m_i^2|_{\rm p}(Q) + \xi \Delta_i^2(Q) 
\label{m2_exp}
\end{eqnarray}
is also a solution if the $\Delta_i^2$ obey the following linear and
homogeneous equation:
\begin{eqnarray}
	\frac{d\Delta_i^2}{d \ln Q} \sim \frac{1}{16\pi^2}
	\sum_j y_j^2 \Delta_j^2 \ .
\label{ddm^2/dt}
\end{eqnarray}
The evolution of the $\Delta_i^2$ depends only on the $\Delta_i^2$,
and the $\Delta_i^2$ are themselves solutions to the RG equations in
the limit $M_a, A_j \rightarrow 0$.

With a given boundary condition, $\Delta_i^2$ may vanish for some $i$
at some renormalization scale $Q_{\rm F}^{(i)}$.  At this scale,
$m_i^2$ is given by $m_i^2|_{\rm p}$ irrespective of $\xi$, and the
family of boundary conditions parameterized by $m_i^2(Q_0)=m_i^2|_{\rm
p}(Q_0)+\xi\Delta_i^2(Q_0)$, with various $\xi$, all yield the same
value of $m_i^2$ at the scale $Q_{\rm F}^{(i)}$.  (Here, $Q_0$ is the
scale where the boundary condition is given.) We call $Q_{\rm
F}^{(i)}$ the focus point scale or ``focus point'' for $m_i^2$.  For
large $\xi$, $m_i^2(Q_0)\gg m_i^2|_{\rm p}(Q_0)$, and if $M_a^2(Q_0),
A^2(Q_0)\sim m_i^2|_{\rm p}(Q_0)$, a large hierarchy $m_i^2(Q_0)\gg
m_i^2(Q_{\rm F}^{(i)})$ is generated.

This concludes our general discussion of focus points.  We now
consider the existence of focus points in the minimal supersymmetric
standard model. Before showing the results of detailed numerical
calculations, we first analyze the focus point behavior by using
one-loop RG equations.  In the minimal supersymmetric standard model,
the possibly large Yukawa couplings are those for the third generation
quarks, $y_t$ and $y_b$.\footnote{For large $\tan\beta$, the $\tau$
Yukawa coupling $y_\tau$ may be sizable, too.  Here, for our
simplified discussion, we neglect $y_\tau$, although its effects are
included in our numerical calculations.}  The system of RG equations
for the $\Delta_i^2$ is then
\begin{eqnarray}
	\frac{d}{d\ln Q}
	\left[ \begin{array}{c}
	\Delta_{H_u}^2 \\
	\Delta_{U_3}^2 \\
	\Delta_{Q_3}^2 \\
	\Delta_{D_3}^2 \\
	\Delta_{H_d}^2
	\end{array} \right]
	= \frac{y_t^2}{8\pi^2}
	\left[ \begin{array}{ccccc}
	3 & 3 & 3 & 0 & 0 \\
	2 & 2 & 2 & 0 & 0 \\
	1 & 1 & 1 & 0 & 0 \\
	0 & 0 & 0 & 0 & 0 \\
	0 & 0 & 0 & 0 & 0
	\end{array} \right]
	\left[ \begin{array}{c}
	\Delta_{H_u}^2 \\
	\Delta_{U_3}^2 \\
	\Delta_{Q_3}^2 \\
	\Delta_{D_3}^2 \\
	\Delta_{H_d}^2
	\end{array} \right]
	+ \frac{y_b^2}{8\pi^2}
	\left[ \begin{array}{ccccc}
	0 & 0 & 0 & 0 & 0 \\
	0 & 0 & 0 & 0 & 0 \\
	0 & 0 & 1 & 1 & 1 \\
	0 & 0 & 2 & 2 & 2 \\
	0 & 0 & 3 & 3 & 3
	\end{array} \right]
	\left[ \begin{array}{c}
	\Delta_{H_u}^2 \\
	\Delta_{U_3}^2 \\
	\Delta_{Q_3}^2 \\
	\Delta_{D_3}^2 \\
	\Delta_{H_d}^2
	\end{array} \right] \ ,
\end{eqnarray}
where $Q_3$, $U_3$, and $D_3$ represent the third generation SU(2)$_L$
doublet, singlet up-type, and singlet down-type squarks, and
$H_u$ and $H_d$ are the up- and down-type Higgses, respectively.  All
other $\Delta_i^2$'s are not coupled to large Yukawa coupling
constants and hence are (almost) scale independent.

To find the focus point, it is simplest to begin by considering small
or moderate values of $\tan\beta$, for which $y_b$ is negligible.  In
this case, $\Delta_{D_3}^2$ and $\Delta_{H_d}^2$ remain constant, but
the RG equations for $\Delta_{H_u}^2$, $\Delta_{U_3}^2$, and
$\Delta_{Q_3}^2$ are solved by
\begin{eqnarray}
	\left[ \begin{array}{c}
	\Delta_{H_u}^2(Q) \\
	\Delta_{U_3}^2(Q) \\
	\Delta_{Q_3}^2(Q)
	\end{array} \right] =
	\kappa_6 \left[ \begin{array}{c}
	3 \\ 2 \\ 1
	\end{array} \right] e^{6I(Q)}
	+ \kappa_0 \left[ \begin{array}{c}
	1 \\ 0 \\ -1
	\end{array} \right]
	+ \kappa_0' \left[ \begin{array}{c}
	0 \\ 1 \\ -1
	\end{array} \right] \ ,
\label{Delta_yt}
\end{eqnarray}
where
\begin{eqnarray}
	I(Q) \equiv \int_{\ln Q_0}^{\ln Q}
	\frac{y_t^2(Q')}{8\pi^2} d\ln Q' \ .
\end{eqnarray}
The $\kappa$'s are constants determined by the boundary conditions at
the scale $Q_0$ and are independent of the renormalization scale $Q$.
For given $\kappa$'s, the focus point for $m_{H_u}^2$ is given by
\begin{eqnarray}
	3 \kappa_6 e^{6I(Q_{\rm F}^{(H_u)})} + \kappa_0 = 0 \ .
\end{eqnarray}

For large $\tan\beta$, we cannot neglect $y_b$, and the above
arguments do not apply.  For a general large $\tan\beta$, the
evolution of the parameters is complicated and will be studied
numerically below.  However, for the specific case $\tan\beta\simeq
m_t/m_b$, we can assume $y_b=y_t$ and follow an analysis similar to
the one above.\footnote{In fact, $y_t$ and $y_b$, even if initially
identical, will be slightly split in their RG evolution by $y_\tau$
and the U(1)$_Y$ gauge interaction.  In this discussion, we neglect
this difference.  This approximation is justified by the numerical
calculations to follow.}  In this case, the $\Delta_i^2$'s evolve
according to
\begin{eqnarray}
	\left[ \begin{array}{c}
	\Delta_{H_u}^2(Q) \\
	\Delta_{U_3}^2(Q) \\
	\Delta_{Q_3}^2(Q) \\
	\Delta_{D_3}^2(Q) \\
	\Delta_{H_d}^2(Q)
	\end{array} \right] =
	\kappa_7 \! \left[ \begin{array}{c}
	3 \\ 2 \\ 2 \\ 2  \\ 3
	\end{array} \right] e^{7I(Q)}
	+ \kappa_5 \! \left[ \begin{array}{c}
	3 \\ 2 \\ 0 \\ -2 \\ -3
	\end{array} \right] e^{5I(Q)}
	+ \kappa_0 \! \left[ \begin{array}{c}
	1 \\ -1 \\ 0 \\ 0 \\ 0
	\end{array} \right]
	+ \kappa_0' \! \left[ \begin{array}{c}
	0 \\ 1 \\ -1 \\ 1 \\ 0
	\end{array} \right]
	+ \kappa_0'' \! \left[ \begin{array}{c}
	0 \\ 0 \\ 0 \\ -1 \\ 1
	\end{array} \right] .
\label{Delta_ytyb}
\end{eqnarray}
The focus point for $m_{H_u}^2$ is then given by
\begin{eqnarray}
	3 \kappa_7 e^{7I(Q_{\rm F}^{(H_u)})} 
	+ 3 \kappa_5 e^{5I(Q_{\rm F}^{(H_u)})} + \kappa_0 = 0 \ .
\end{eqnarray}

The actual focus point depends on the boundary condition.  Here, we
first consider the well-studied case of minimal supergravity.  In this
framework, the supersymmetric Lagrangian is specified by five new
fundamental parameters: the universal scalar mass $m_0$, the unified
gaugino mass $M_{1/2}$, the supersymmetric Higgs mass $\mu_0$, the
universal trilinear coupling $A_0$, and the bilinear Higgs scalar
coupling $B_0$.  These parameters are given at the GUT scale $M_{\rm
GUT}$, which, in our analysis, is defined as the scale where the
SU(2)$_L$ and U(1)$_Y$ gauge couplings meet.  (Numerically, $M_{\rm
GUT}\simeq 2\times 10^{16}~{\rm GeV}$.)  All the supersymmetry
breaking scalar masses are universal at $M_{\rm GUT}$, and we may
take\footnote{We choose $m_i^2|_{\rm p} (M_{\rm GUT})=0$ so that
$m_i^2|_{\rm p}$ is independent of $m_0$ and is always ${\cal
O}(M_{1/2}^2)$ (or ${\cal O}(A_0^2)$) or smaller.}
\begin{eqnarray}
	m_i^2|_{\rm p} (M_{\rm GUT}) &=& 0 \ ,
\\
	\xi \Delta_i^2 (M_{\rm GUT}) &=& m_0^2 \ .
\end{eqnarray}

With these boundary conditions, the coefficients $\kappa$ are

\begin{eqnarray}
(\kappa_6, \kappa_0, \kappa_0') &=& m_0^2 \left( \frac{1}{2},
-\frac{1}{2}, 0 \right) \ , \quad y_b \ll y_t \ , \\
(\kappa_7,\kappa_5,\kappa_0,\kappa_0',\kappa_0'') &=&
m_0^2 \left( \frac{3}{7}, 0, -\frac{2}{7}, -\frac{1}{7}, 
-\frac{2}{7} \right) \ , \quad y_b = y_t \ ,
\end{eqnarray}
and the focus point scale is determined by

\begin{eqnarray}
e^{6I(Q)} &=& 1/3 \ , \ {\rm for}~y_b\ll y_t \ , \label{fp1} \\
e^{7I(Q)} &=& 2/9 \ , \ {\rm for}~y_b=y_t  \label{fp2} \ .
\end{eqnarray}
Note that the $I(Q)$ in Eqs.~(\ref{fp1}) and (\ref{fp2}) are not
identical, as $y_t$ runs differently for small and large $\tb$.

Eqs.~(\ref{fp1}) and (\ref{fp2}) determine the focus point scale in
terms of the gauge couplings and the top quark Yukawa coupling, or
equivalently, the top quark mass.  Remarkably, for the physical gauge
couplings and top quark mass $\mt \approx 174~{\rm GeV}$, both
conditions yield focus points that are very close to the weak scale!
Thus, in minimal supergravity, the weak scale value of $m_{H_u}^2$ is
highly insensitive to the universal scalar mass $m_0$.

We now show numerically that the focus point is near the weak scale
for $\mt \approx 174 \gev $. (For an analytical discussion, see the
Appendix.)  To study the focus point more carefully, we have evolved
the supersymmetry breaking parameters with the full two-loop RG
equations~\cite{2loop RGEs}.  The one-loop threshold corrections from
supersymmetric particles to the gauge and Yukawa coupling constants
are also included~\cite{BMP,PBMZ}.  We take as inputs $\alpha_{\rm
em}^{-1} = 137.0359895$, $G_F = 1.16639\times 10^{-5}$, $\alpha_s(m_Z)
= 0.117$, $m_Z = 91.187~{\rm GeV}$, $m_\tau^{DR}(m_Z) = 1.7463~{\rm
GeV}$, bottom quark pole mass $m_b=4.9~{\rm GeV}$, and, unless
otherwise noted, top quark pole mass $m_t=174~{\rm GeV}$.

\begin{figure}[tbp]
\postscript{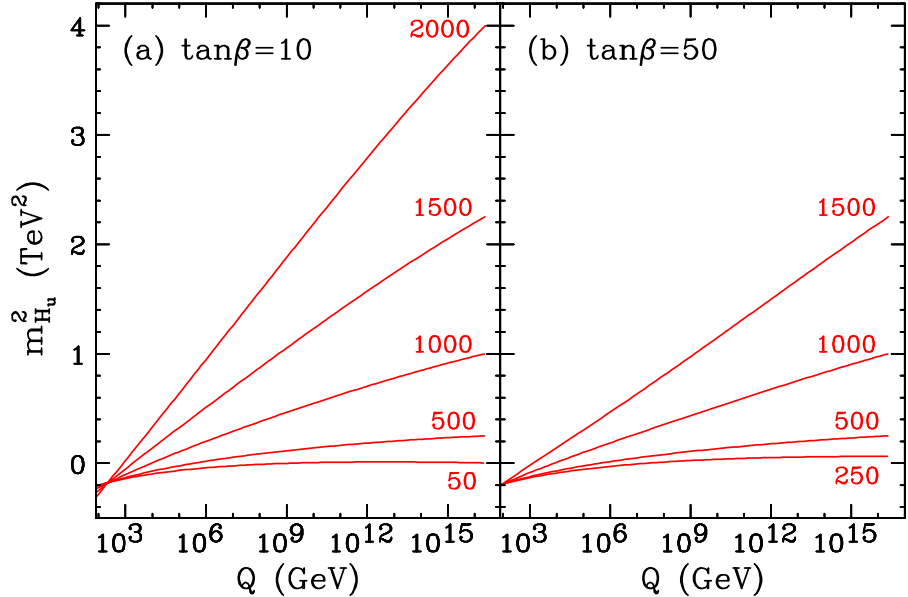}{0.74}
\caption
{The RG evolution of $m_{H_u}^2$ for (a) $\tb=10$ and (b) $\tb=50$,
several values of $m_0$ (shown, in GeV), $\mgaugino = 300$ GeV,
$A_0=0$, and $\mt = 174$ GeV.  For both values of $\tb$, $m_{H_u}^2$
exhibits an RG focus point near the weak scale, where $Q_{\rm
F}^{(H_u)}\sim {\cal O}(100~{\rm GeV})$, irrespective of $m_0$. }
\label{fig:run}
\end{figure}

The scale dependence of $m_{H_u}^2$ for various values of $m_0$ in
minimal supergravity is shown in Fig.~\ref{fig:run}.  To high
accuracy, all of the RG trajectories meet at $Q\sim {\cal O}(100~{\rm
GeV})$.  In fact, in this case, the weak value of $m_{H_u}^2$ is
determined by the other fundamental parameters $M_{1/2}$ and $A_0$,
and hence at least one of these parameters is required to be ${\cal
O}(100~{\rm GeV})$.

\begin{figure}[tbp]
\postscript{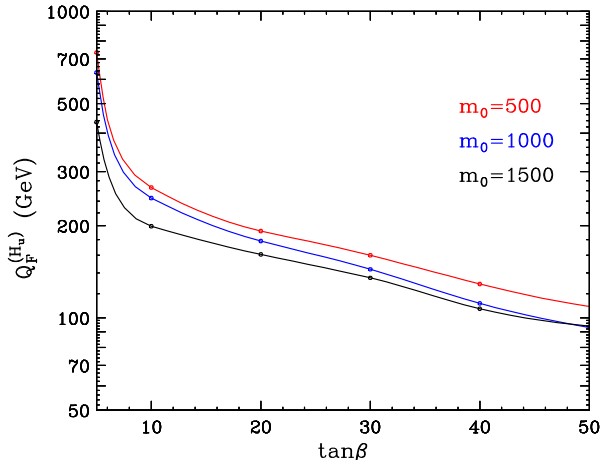}{0.49}
\caption
{The focus point renormalization scale $Q_{\rm F}^{(H_u)}$ as a
function of $\tb$ for $m_0 = 500$, 1000, and 1500 GeV (from above),
$\mgaugino = 300$ GeV, $A_0=0$, and $\mt = 174$ GeV. The focus point
scale is defined as the scale at which $\partial m_{H_u}^2 / \partial
m_0 = 0$. }
\label{fig:focus}
\end{figure}

In Fig.~\ref{fig:run}, two values of $\tb$ were presented.  In
Fig.~\ref{fig:focus}, we show the focus point scale of $m_{H_u}^2$ as
a function of $\tan\beta$.  The focus point is defined here as the
scale where $\partial m_{H_u}^2/\partial m_0=0$.  As noted above, we
have included the low-energy threshold corrections to the gauge and
Yukawa coupling constants, which depend on the soft supersymmetry
breaking parameters.  As a result, the RG trajectories do not all meet
at one scale, and the focus point given in Fig.~\ref{fig:focus} has a
slight dependence on $m_0$.  For small values of $\tb$, say, $\tb \sim
2-3$, the focus point is at very large scales.  However, the important
point is that, for all values of $\tb \agt 5$, including both moderate
values of $\tb$ and large values where $y_b$ and $y_\tau$ are not
negligible, $Q_{\rm F}^{(H_u)}\sim {\cal O}(100~{\rm GeV})$, and the
weak scale value of $m_{H_u}^2$ is insensitive to $m_0$.

So far, we have considered only the case of a universal scalar mass.
However, the $m_{H_u}^2$ focus point remains at the weak scale
for a much wider class of boundary conditions.  For example, for small
$\tan\beta$, Eq.~(\ref{Delta_yt}) shows that the parameter $\kappa_0'$
does not affect the evolution of $m_{H_u}^2$.  As a result, the focus
point of $m_{H_u}^2$ does not change even if we vary $\kappa_0'$, and
the weak scale focus point is realized with any boundary
condition of the form

\begin{equation}
(m_{H_u}^2, m_{U_3}^2, m_{Q_3}^2)\propto (1, 1+x, 1-x) \ ,
\label{generalsmalltb}
\end{equation}
with $x$ an arbitrary constant.  Similarly, for the case of $y_b=y_t$,
the possible variation is
\begin{equation}
(m_{H_u}^2, m_{U_3}^2, m_{Q_3}^2, m_{D_3}^2,
m_{H_d}^2) \propto(1,1+x,1-x,1+x-x',1+x') \ ,
\label{generallargetb}
\end{equation}
with both $x$ and $x'$ arbitrary.

Another possible modification of the boundary conditions may be seen
by viewing the $m_i^2|_{\rm p}(Q_0)$ as perturbations.  Since it was
never necessary to specify the particular solution in the general
focus point analysis, we may consider arbitrary $m_i^2|_{\rm p}(Q_0)$
without changing the focus point scale.  The only constraint on
$m_i^2|_{\rm p}$ is from naturalness.  As will be discussed in the
next section, $m_{H_u}^2$ is required to be ${\cal O}((100~{\rm
GeV})^2)$ at the weak scale for natural electroweak symmetry breaking.
As a result, $m_{H_u}^2|_{\rm p}$, $m_{U_3}^2|_{\rm p}$, and
$m_{Q_3}^2|_{\rm p}$ (and also $m_{H_d}^2|_{\rm p}$ and
$m_{D_3}^2|_{\rm p}$ for large $\tan\beta$) are required to be of the
order of the weak scale.  Therefore, deviations from the boundary
conditions of Eqs.~(\ref{generalsmalltb}) and (\ref{generallargetb})
of order $\delta m^2 \sim {\cal O}((100~{\rm GeV})^2)$ are acceptable
and do not lead to fine-tuning problems.  Similar arguments show that
deviations $M_a, A_j \sim {\cal O}(100~{\rm GeV})$ are allowed.  In
particular, the focus point is independent of gaugino mass or $A$
parameter universality, and may therefore be found in many other
frameworks.  For example, focus points also exist in anomaly-mediated
supersymmetry breaking models with additional universal scalar
masses~\cite{anom}.

Before closing this section, we discuss another way of formulating the
focus point, which was originally used in Ref.~\cite{barbieri} for the
specific case of minimal supergravity. By dimensional analysis, the
evolution of $m_{H_u}^2$ may be parameterized as
\begin{eqnarray}
	m_{H_u}^2 (Q) = \eta_{m_0^2} (Q) m_0^2
	+ \eta_{M_{1/2}^2} (Q) M_{1/2}^2
	+ \eta_{M_{1/2}A_0} (Q) M_{1/2}A_0
	+ \eta_{A_0^2} (Q) A_0^2 \ ,
\label{m^2(eta)}
\end{eqnarray}
where the coefficients $\eta$ are determined by the (dimensionless)
gauge and Yukawa coupling constants, and are independent of the
(dimensionful) supersymmetry breaking parameters. In this formulation,
the focus point is given by the scale where $\eta_{m_0^2}=0$, since at
that scale, the value of $m_{H_u}^2$ is insensitive to
$m_0$.\footnote{Notice that in this formulation, it is clear that all
RG trajectories meet at a focus point to all orders in the RG
equations.  Of course, as noted above, threshold effects smear out the
focus point slightly --- see Fig.~\ref{fig:focus}. }  Based on this
observation, it was noted in Ref.~\cite{barbieri} that, for minimal
supergravity, and neglecting $y_b$, $\eta_{m_0^2} = 0$ at the weak
scale, and the weak scale becomes very insensitive to the universal
scalar mass $m_0$, for $m_t\simeq 160-170~{\rm GeV}$.  As may be seen
from the general analysis of focus points above, however, this
conclusion holds much more generally: it is valid even when $y_b$ is
not negligible, holds for the more general boundary conditions given
in Eqs.~(\ref{generalsmalltb}) and (\ref{generallargetb}), and is
independent of all other scalar masses.  In addition, as noted above
and as is evident from Eq.~(\ref{m^2(eta)}), the conclusion is valid
also for non-universal gaugino masses and $A$ parameters, as long as
they are not too much larger than the weak scale.

\section{Naturalness}
\label{sec:natural}

In the previous section, we saw that the weak scale value of
$m_{H_u}^2$ is highly insensitive to the high scale scalar mass
boundary conditions in a class of models that includes minimal
supergravity.  This fact has important implications for the
naturalness of the gauge hierarchy, since $m_{H_u}^2$ determines, to a
large extent, the shape of the Higgs potential.

In minimal supergravity, $m_{H_u}^2$ and other sfermion masses have
the same origin, the universal scalar mass $m_0$, and it has typically
been believed that naturalness constraints on $m_{H_u}^2$ also give
similar bounds on the sfermion masses.  Such beliefs have led to great
optimism in the search for scalar superpartners at future colliders in
the framework of minimal supergravity~\cite{Snowmass}.  However, as we
have seen, the relation between $m_{H_u}^2$ and other sfermion masses
is not as trivial as typically assumed.  In the following, we
therefore reconsider the naturalness bounds on the sfermion masses in
the minimal supergravity model.

To begin, it is instructive to start with the tree-level expression
for the weak scale.  The $Z$ boson mass is determined by minimizing
the tree-level Higgs potential to be
\begin{eqnarray}
	\frac{1}{2} m_Z^2 =
	\frac{m_{H_d}^2-m_{H_u}^2 \tan^2\beta}{\tan^2\beta -1}
	- \mu^2 \ .
\label{mZ^2/2}
\end{eqnarray}
For all $\tb$, $m_{H_u}^2 \gg m_Z^2$ is disfavored by the naturalness
criterion, as in that case, a large cancellation between $m_{H_u}^2$
and $\mu^2$ is needed to arrive at the physical value of the weak
scale.  For moderate and large values of $\tb$, however, $m_{H_d}^2
\gg m_Z^2$ does not necessarily lead to fine-tuning.

For more detailed discussions of naturalness, it is convenient to
define a quantitative measure of
fine-tuning~\cite{barbieri,natural1,natural1.5,Anderson,natural2}.
Following previous work, we use the sensitivity of the weak scale
(i.e., $m_Z$) to fractional variations in the fundamental parameters
as such a measure.

In any discussion of naturalness, several subjective choices must be
made.  The choice of supersymmetry breaking framework is crucial.  For
example, in GUT models, the gaugino masses are all governed by one
parameter, whereas in general, all three gaugino masses may be varied
independently, and the sensitivity of the weak scale to each of them
must be considered.  In the following, we will specialize to minimal
supergravity.  As noted previously, minimal supergravity introduces
five new fundamental parameters: $m_0$, $M_{1/2}$, $\mu_0$, $A_0$, and
$B_0$.  All quantities at the weak scale are fixed by these
parameters.  In particular, the vacuum expectation values of the Higgs
bosons depend on these quantities.  Therefore, some combination of
them is constrained to yield the correct $Z$ boson mass.  (At tree
level, this constraint is that of Eq.~(\ref{mZ^2/2}). At one-loop, the
Higgs masses squared are shifted by the corresponding tadpole
contributions \cite{EP 1-loop}, as will be discussed below.)  {}From
the low energy point of view, it is therefore more convenient to
consider $\tb$ and $\sign(\mu)$ as free parameters, instead of $\mu_0$
and $B_0$.

We adopt the following procedure to calculate the magnitude of
fine-tuning at all physically viable parameter points:

\noindent
({\it i}) We consider the minimal supergravity framework with its
$4+1$ input parameters
\begin{eqnarray}
	\{ P_{\rm input} \} =
	\{ m_0, M_{1/2}, A_0, \tan\beta, \sign(\mu) \} \ .
\end{eqnarray}
Any point in the parameter space of minimal supergravity is specified
by these parameters.

\noindent
({\it ii}) The naturalness of each point is then calculated by first
determining all the parameters of the theory (Yukawa couplings, soft
supersymmetry breaking masses, etc.), consistent with low energy
constraints.  RG equations are used to relate high and low energy
boundary conditions.  In particular, using the relevant radiative
breaking condition, $|\mu_0|$ and $B_0$ are determined consistent with
the low energy constraints.

\noindent
({\it iii}) We choose to consider the following set of (GUT scale)
parameters to be free, continuously valued, independent, and
fundamental: 
\begin{equation}
\{ a_i\} =\{ m_0, \mgaugino, \mu_0, A_0, B_0 \} \ .
\label{a_i}
\end{equation}

\noindent
({\it iv}) All observables, including the $Z$ boson mass, are then
reinterpreted as functions of the fundamental parameters $a_i$, and
the sensitivity of the weak scale to small fractional variations in
these parameters is measured by the sensitivity coefficients
\begin{eqnarray}
	c_a \equiv
	\left| \frac{\partial \ln m_Z^2}{\partial \ln a} \right| \ ,
\label{ca}
\end{eqnarray}
where all other fundamental (not input) parameters are held fixed in
the partial derivative.\footnote{The sensitivity of $v^2=v_u^2+
v_d^2$, where $v_u$ and $v_d$ are the vacuum expectation values of the
up- and down-type Higgs scalars, respectively, may be a more accurate
choice, especially if variations of the gauge coupling constants are
considered.  In this paper, however, we follow the literature and
define sensitivity coefficients as in Eq.~(\ref{ca}).}

\noindent
({\it v}) Finally, we form the fine-tuning parameter
\begin{eqnarray}
	c \equiv {\rm max} \{ c_{m_0}, c_{M_{1/2}}, c_{\mu_0},
	c_{A_0}, c_{B_0} \} \ ,
\end{eqnarray}
which is taken as a measure of the naturalness of point $\{ P_{\rm
input} \}$, with large $c$ corresponding to large fine-tuning.

Among the choices made in the prescription above, the choice of
fundamental parameters $a_i$ is of particular importance. This choice
varies throughout the
literature~\cite{barbieri,natural1,natural1.5,Anderson,natural2}.
Since we are interested in the naturalness of the supersymmetric
solution to the gauge hierarchy problem, we find it most reasonable to
include only supersymmetry breaking parameters (and $\mu$, as its
origin is likely to be tied to supersymmetry breaking) and not
standard model parameters, such as $y_t$ or the strong coupling.  We
will return to this issue in Sec.~\ref{sec:summary}.

Given the prescription for defining fine-tuning described above, we
may now present the numerical results.  In minimizing the Higgs
potential, we use the one-loop corrected Higgs potential, calculated
with parameters evaluated with two-loop RG equations. Denoting the
physical stop masses by $m_{\tilde t_1}$ and $m_{\tilde t_2}$, we
choose to minimize the potential at the scale
\begin{eqnarray}
Q_{\tilde t} = (m_{\tilde t_1}m_{\tilde t_2})^{1/2} \ ,
\end{eqnarray}
where the one-loop corrections to the Higgs potential tend to be
smallest~\cite{natural1.5}.  In terms of the fundamental parameters,
$Q_{\tilde t}\simeq 0.5 (m_0^2+4M_{1/2}^2)^{1/2}$.

\begin{figure}[tbp]
\postscript{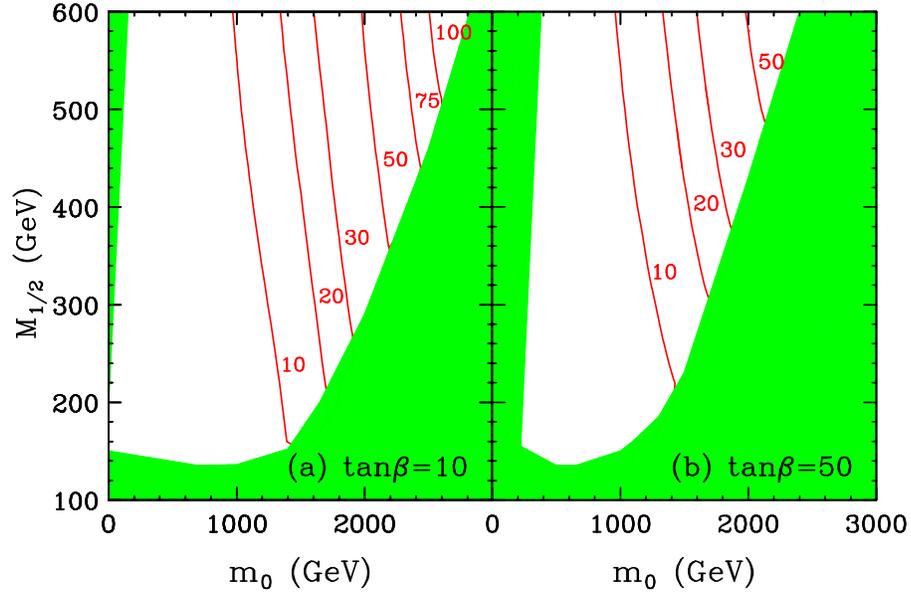}{0.74}
\caption
{Contours of constant sensitivity parameter $c_{m_0}$ in the
$(m_0, \mgaugino)$ plane for (a) $\tb = 10$ and (b) $\tb = 50$,
$A_0=0$, $\mu>0$, and $\mt = 174$ GeV. The bottom and right shaded
region is excluded by the chargino mass limit of 90 GeV.  The top left
region is also excluded if a neutral LSP is required. }
\label{fig:c_m0}
\end{figure}

\begin{figure}[tbp]
\postscript{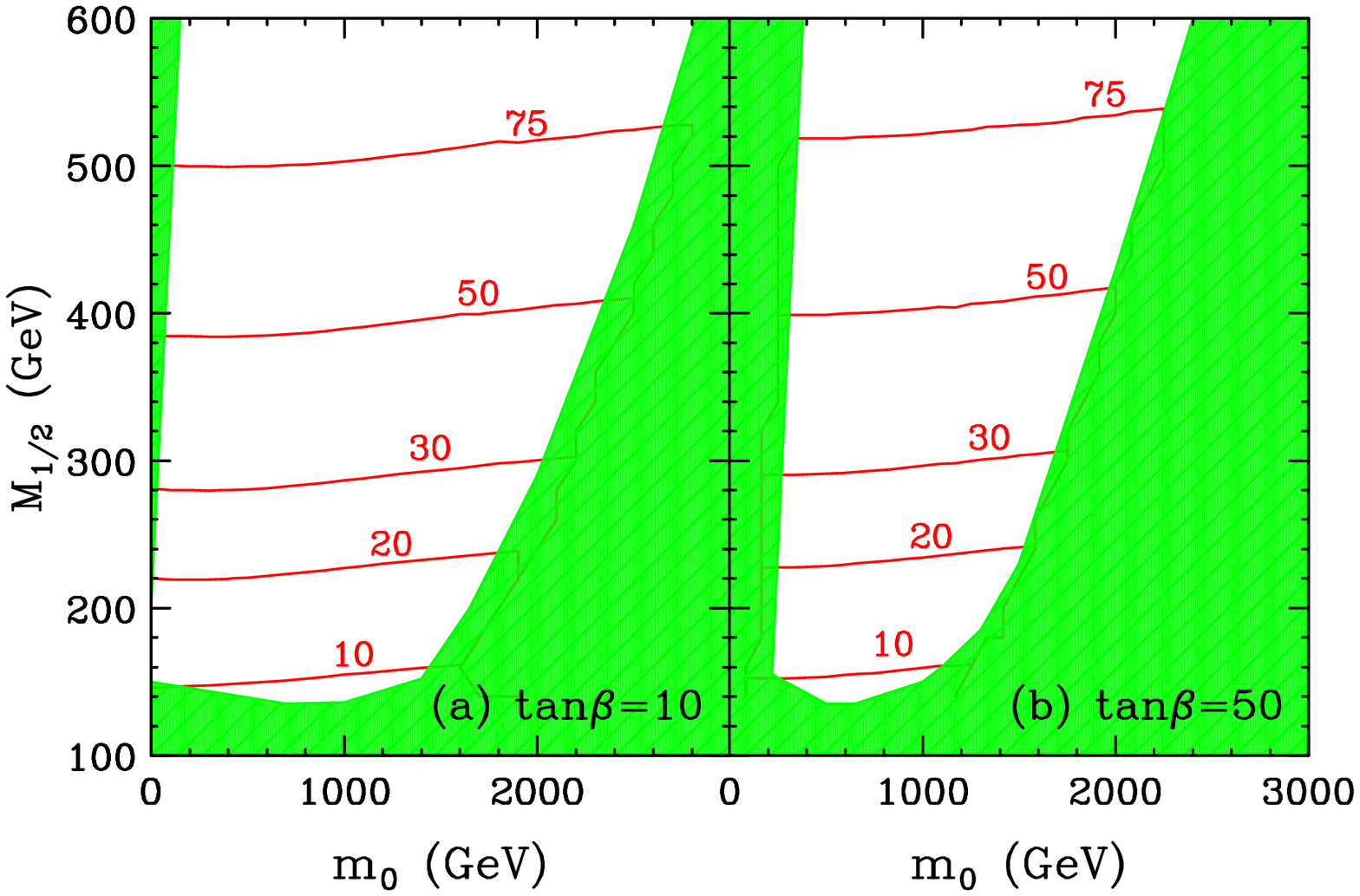}{0.74}
\caption
{As in Fig.~\ref{fig:c_m0}, but for the sensitivity parameter
$c_{\mgaugino}$. }
\label{fig:c_mgaugino}
\end{figure}

\begin{figure}[tbp]
\postscript{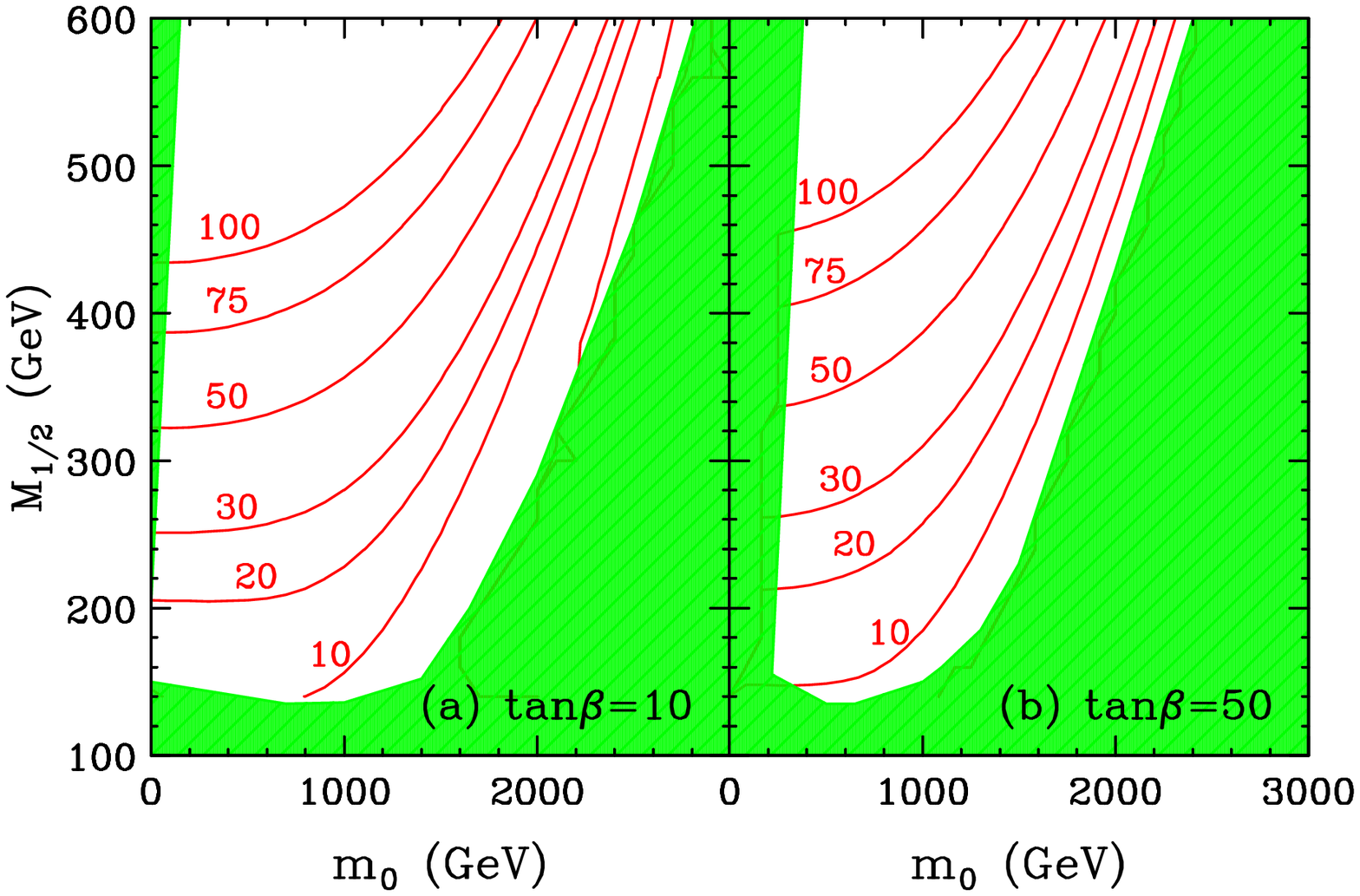}{0.74}
\caption
{As in Fig.~\ref{fig:c_m0}, but for the sensitivity parameter
$c_{\mu_0}$. }
\label{fig:c_mu0}
\end{figure}

In Figs.~\ref{fig:c_m0}--\ref{fig:c_mu0}, we give contours of constant
$c_{m_0}$, $c_{M_{1/2}}$, and $c_{\mu_0}$ in the $(m_0, M_{1/2})$
plane for $\tan\beta =10$ and 50.  (The parameters $c_{A_0}$ and
$c_{B_0}$ are typically negligible relative to these, and are so for
the parameter ranges displayed.)

Several features of these figures are noteworthy.  First, as is
evident from Fig.~\ref{fig:c_m0}, the fine-tuning coefficient
$c_{m_0}$ is small even for scalar masses as large as $m_0 \sim 2
\tev$.  This is a consequence of the focus point behavior of
$m_{H_u}^2$: for moderate and large $\tb$, $m_{H_u}^2$ is insensitive
to $m_0$, and, therefore, so is $m_Z^2$.  The deviation of $c_{m_0}$
from zero for very large $m_0$ is a consequence of the fact that the
focus point does not coincide exactly with the electroweak scale, or,
more precisely, with $Q_{\tilde{t}}$. We will explain this statement
in more detail below.

On the other hand, large gaugino masses lead to large $m_{H_u}^2$
through RG evolution, and hence $c_{M_{1/2}}$ increases as $M_{1/2}$
increases, as shown in Fig.~\ref{fig:c_mgaugino}.  Multi-TeV values of
$\mgaugino$ are therefore inconsistent with naturalness. 

The behavior of $c_{\mu_0}$, presented in Fig.~\ref{fig:c_mu0}, is
also interesting.  Since $\mu \propto \mu_0$, Eq.~(\ref{mZ^2/2})
implies $c_{\mu_0}\approx 4\mu^2/m_Z^2$.  In particular, $c_{\mu_0}$
is small in the region bordering the excluded right-hand region, as
there $\mu$ is suppressed by a cancellation between the $\eta_{m_0^2}$
and $\eta_{M_{1/2}^2}$ terms in Eq.~(\ref{m^2(eta)}). However, in this
region, $c_{m_0}$ and $c_{M_{1/2}}$ are large and this region is
fine-tuned; the simple criterion of requiring low $\mu$ for
naturalness~\cite{mu criterion} fails here.

\begin{figure}[tbp]
\postscript{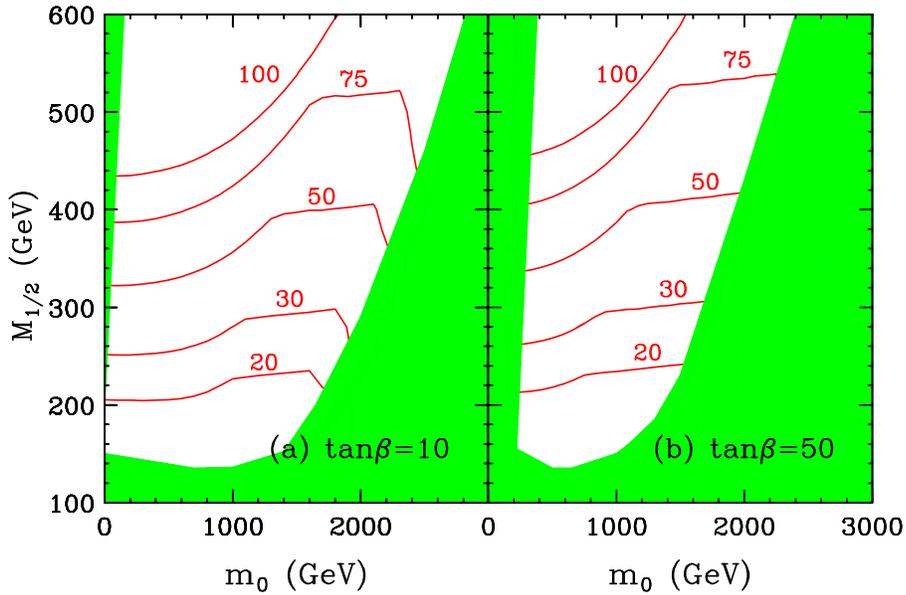}{0.74}
\caption
{As in Fig.~\ref{fig:c_m0}, but for the fine-tuning parameter $c$.}
\label{fig:m0M1/2}
\end{figure}

In Fig.~\ref{fig:m0M1/2}, we show the overall fine-tuning parameter
$c$, the maximum of $c_a$, in the $(m_0, M_{1/2})$ plane.  The
fine-tuning $c$ is determined by $c_{\mu_0}$, $c_{M_{1/2}}$, and
$c_{m_0}$.  For small $m_0$, $c_{\mu_0}$ is the largest.  As $m_0$
increases, however, $c_{M_{1/2}}$ becomes dominant, and $c$ is
therefore almost independent of $m_0$ in this region.  Finally, for
extremely large $m_0$, $c_{m_0}$ becomes important.  (For large
$\tan\beta$, large $m_0$ is excluded by the chargino mass limit before
$c_{m_0}$ becomes dominant.  As a result, we do not see the $c_{m_0}$
segment in Fig.~\ref{fig:m0M1/2}b.) Note that, for fixed $\mgaugino$
in Fig.~\ref{fig:m0M1/2}, values of $m_0 \agt 1 \tev$ are actually
{\em more} natural than small $m_0$: for large $m_0$, the parameter
$\mu$, and therefore $c_{\mu_0}$, is reduced.  Of course, eventually
as $m_0$ increases to extremely large values, either $\mu$ becomes so
small that the chargino mass bound is violated or $c_{m_0}$ becomes
large, and so very large $m_0$ is either excluded or disfavored.

As one can see, regions of parameter space with $m_0\sim 2-3~{\rm
TeV}$ are as natural as regions with $m_0\lesssim 1~{\rm TeV}$. As
will be discussed more fully in Sec.~\ref{sec:implications}, in the
region of parameter space with $m_0\sim 2-3~{\rm TeV}$, all squarks
and sleptons have multi-TeV masses, and discovery of these scalars
will be extremely challenging at near future colliders. On the other
hand, the gaugino mass $M_{1/2}$ cannot be multi-TeV since it
generates unnaturally large $m_{H_u}^2$.  Although the focus point
mechanism allows multi-TeV scalars consistent with naturalness, the
same conclusion does not apply to gauginos and Higgsinos.

\begin{figure}[tbp]
\postscript{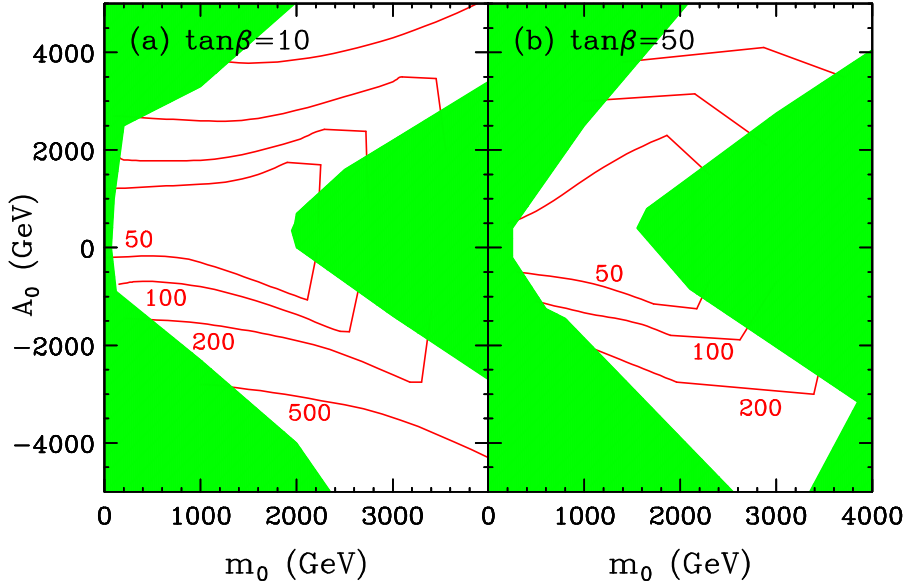}{0.74}
\caption
{Contours of constant fine-tuning $c$ in the $(m_0, A_0)$ plane for
(a) $\tb = 10$ and (b) $\tb = 50$, $\mgaugino = 300$ GeV, $\mu>0$, and
$\mt = 174$ GeV.  The shaded regions in the upper and lower left
corners are excluded by top squark mass bounds.  The shaded region on
the right is excluded by the chargino mass limit, while the region in
the lower right corner of panel (b) is excluded by Higgs searches, in
particular, searches for the CP-odd Higgs boson.  The thin strip on
the left in panel (a) is excluded if a neutral LSP is required.}
\label{fig:m0A0}
\end{figure}

As discussed in Sec.~\ref{sec:natural}, we expect also that the $A$
parameters should be bounded by naturalness to be near the weak scale.
In Fig.~\ref{fig:m0A0}, we present contours of constant $c$ in the
$(m_0, A_0)$ plane.  As expected, large $A$ terms lead to large
$m_{H_u}^2$, and $A_0$ is also required to be ${\cal O}(100~{\rm
GeV})$.  In Fig.~\ref{fig:m0A0}, for increasing $m_0$, $c$ is
determined successively by $c_{\mu_0}$, $c_{A_0}$ and $c_{m_0}$. (The
$c_{m_0}$ segments are missing for the $A_0 >0$ contours in
Fig.~\ref{fig:m0A0}b.)

\begin{figure}[tbp]
\postscript{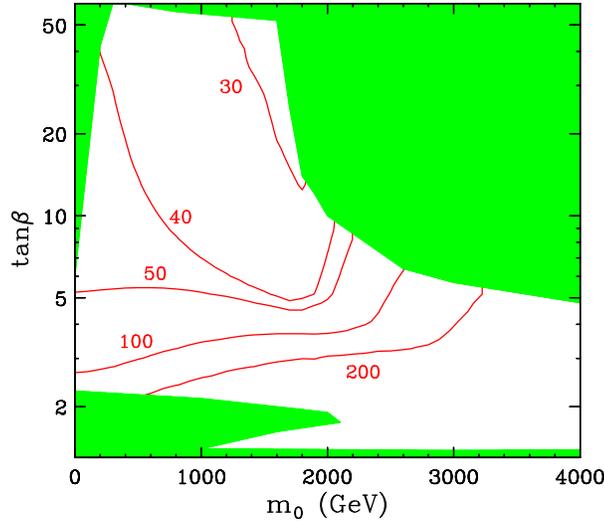}{0.49}
\caption
{Contours of constant fine-tuning $c$ in the $(m_0, \tb)$ plane for
$\mgaugino = 300 \gev$, $A_0=0$, $\mu>0$, and $\mt=174 \gev$. The thin
shaded strip on the bottom is excluded by the requirement that $y_t$
remain perturbative up to the GUT scale ($y_t(\mgut)<3.5$). The region
in the lower left corner is excluded by the LEP Higgs mass limit
$m_h>95$ GeV, the region in the upper right corner is excluded by
chargino searches, and the region at very large $\tan\beta$ on the top
is again ruled out by Higgs searches, as the CP-odd Higgs becomes too
light. Finally, the upper left region is also excluded if a neutral
LSP is required.}
\label{fig:m0tb}
\end{figure}

The dependence of the fine-tuning parameter $c$ on $\tb$ is
illustrated in Fig.~\ref{fig:m0tb}. {}From this figure, for a given
$\tan\beta$ and maximal allowed $c$, we can determine the upper bound
on $m_0$.  The exact range of $c$ required for a natural model is, of
course, subjective.  However, taking as an example the requirement
$c\leq 50$, corresponding to $\mu \alt 300 \gev$ at parameter points
where $c = c_{\mu_0}$, we find $m_0\alt 2~{\rm TeV}$ for $\tan\beta =
10$.

So far, we have assumed $m_t=174~{\rm GeV}$ in our calculations.
However, given the experimentally allowed range $m_t=173.8\pm 5.2~{\rm
GeV}$~\cite{PDG}, we now consider the top quark mass dependence of the
naturalness constraint on $m_0$.  This can be understood only after
accounting for the one-loop corrections to the Higgs effective
potential.\footnote{In fact, the sensitivity coefficient $c_{m_0}$ can
only be reliably calculated, and, formally, is only meaningful, at
one-loop, since at tree-level there is no preferred scale at which to
enforce Eq.~(\ref{mZ^2/2}).} At one-loop, the relation between $m_Z$
and the Higgs mass parameters, Eq.~(\ref{mZ^2/2}), is modified
to~\cite{PBMZ,EP 1-loop}

\begin{eqnarray}
\frac{1}{2} m_Z^2 &=&
\frac{(m_{H_d}^2-T_d/v_d)-(m_{H_u}^2-T_u/v_u)\tan^2\beta}
     {\tan^2\beta -1} - \mu^2 - {\rm Re}\ \Pi^T_{ZZ}(M_Z)\nonumber \\ 
&\simeq& -\ m_{H_u}^2\,+\,T_u/v_u\,-\, \mu^2, \ 
{\rm for}~\tan\beta\gg 1 \ ,
\label{1-loop minim cond}
\end{eqnarray}
where $T_u$ and $T_d$ are the tadpole contributions to the effective
potential and $\Pi^T_{ZZ}(p)$, the transverse part of the $Z$ boson
self-energy, is negligible. In minimal supergravity, the dominant
terms in $T_u$ and $T_d$ are from third generation squark loops and
have the generic form

\begin{equation}
{T_{u,d}\over v_{u,d}}\ \sim\ {3 y_{t,b}^2\over 16\pi^2}\
m^2_{\tilde{t}, \tilde{b}}\ \left[{1\over2}-\ln 
\left({m_{\tilde{t}, \tilde{b}}\over Q}\right)\right] \ .
\label{tadpole}
\end{equation}
Using Eqs.~(\ref{1-loop minim cond}), (\ref{m^2(eta)}) and
(\ref{tadpole}), we find

\begin{eqnarray}
{1\over2}\,m_Z^2&\simeq&-\,\left\{\ \eta_{m_0^2}(Q)\, 
m_0^2 + \ldots\right\}
+\, {3 y_t^2\over 16\pi^2}\ m^2_{\tilde t}\
\left[{1\over2}-\ln\left({m_{\tilde t}\over Q}\right)\right]+\ldots\ 
-\, \mu^2 \nonumber \\
&\equiv& -\, \left[ \eta_{m_0^2}(Q) + \eta'_{m_0^2}(Q)\right]
 m_0^2 - \mu^2\, +\,  \ldots \ ,
\label{eta'}
\end{eqnarray}
where $\eta'_{m_0^2}(Q)$ encodes the dependence on $m_0$ arising at
one-loop through the tadpole.

Recall that the focus point $Q_{\rm F}^{(H_u)}$ is defined by

\begin{equation}
\eta_{m_0^2}(Q=Q_{\rm F}^{(H_u)})\ =\ 0 \ ,
\end{equation}
while empirically we find 

\begin{equation}
\eta'_{m_0^2}(Q=Q_{\tilde t})\ \approx \ 0 \ ,
\end{equation}
which may also be understood to a good approximation from
Eq.~(\ref{eta'}).  We have already seen from Fig.~\ref{fig:focus} that
for $m_t=174$ GeV, $Q_{\rm F}^{(H_u)} \sim {\cal O} (100 \gev)$, well
below the typical stop mass scale $Q_{\tilde t}$.  The sensitivity of
$m_{H_u}^2$ to $m_0$ may then be understood in two ways: either we may
minimize the potential at $Q_{\tilde t}$ where the tadpole
contributions are negligible, but $\eta_{m_0^2}$ is non-vanishing, or
we may choose to minimize the potential at $Q_{\rm F}^{(H_u)}$, in
which case $\eta_{m_0^2} = 0$, but $m_0$ dependence arises from
non-vanishing tadpole contributions. Either way, there is some
residual dependence on $m_0$, as can be seen in Fig.~\ref{fig:c_m0}.

If $Q_{\rm F}^{(H_u)}$ and $Q_{\tilde t}$ are identical, however,
$\eta_{m_0^2}(Q) + \eta'_{m_0^2}(Q)$ vanishes at $Q_{\tilde t}$.  This
can happen in two ways: for a fixed top quark mass, $Q_{\tilde t}$ can
be lowered to $Q_{\rm F}^{(H_u)}$ by lowering $m_0$, or, for fixed
large $m_0$, $Q_{\rm F}^{(H_u)}$ may be raised to $Q_{\tilde t}$ by
increasing $\mt$ (and thereby increasing the top Yukawa
renormalization effect on $m_{H_u}^2$).

\begin{figure}[tbp]
\postscript{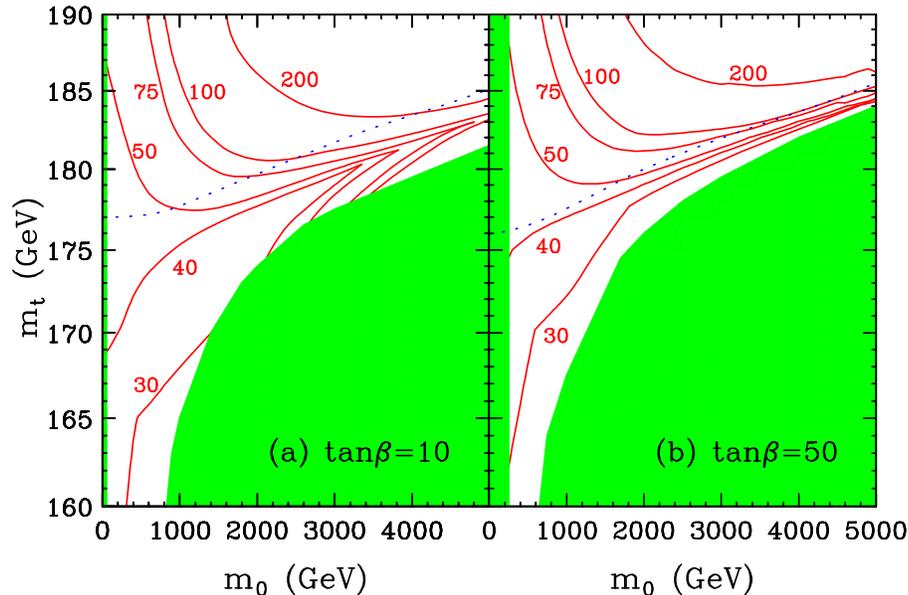}{0.74}
\caption
{Contours of constant fine-tuning $c$ in the $(m_0, \mt)$ plane for
$\mgaugino = 300 \gev$, $A_0=0$, $\tb = 10$, and $\mu>0$. The bottom
and right shaded region is excluded by the chargino mass limit of 90
GeV.  The left region is excluded if a neutral LSP is required.  The
sensitivity coefficient $c_{m_0}$ vanishes on the dotted contour (see
text).}
\label{fig:m0mt}
\end{figure}

In Fig.~\ref{fig:m0mt}, we present contours of fine-tuning $c$ in the
$(m_0, m_t)$ plane.  On the dotted contour, the focus point and
$Q_{\tilde{t}}$ coincide, and $c_{m_0} = 0$.  This occurs for $\mt$
above 174 GeV, in accord with the discussion above.  More generally,
the sensitivity $c_{m_0}$ is indeed reduced for larger $\mt$. As a
result, the upper bound on $m_0$ is increased for larger $\mt$, and
for $\mt \simeq 180 \gev$, near the 1$\sigma$ upper bound, $m_0\sim
3.4~{\rm TeV}$ is allowed for $c\leq 50$.  For smaller top quark mass,
the naturalness bound on $m_0$ becomes more stringent.  Thus, while
the focus point behavior persists for all $\mt$ within current
experimental bounds, future improvements in top mass
measurements~\cite{top quark mass} may provide important information
about the extent to which multi-TeV scalars are allowed in minimal
supergravity.

\begin{figure}[tbp]
\postscript{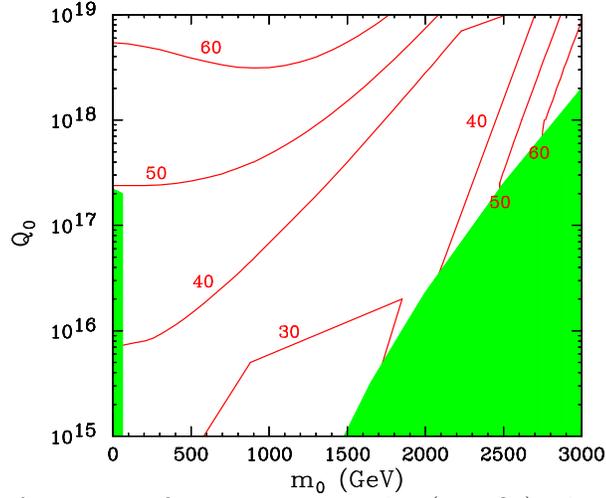}{0.49}
\caption
{Contours of constant fine-tuning $c$ in the $(m_0, \minit)$ plane for
$\mgaugino = 300 \gev$, $A_0=0$, $\tb=10$, $\mu>0$, and $\mt = 174
\gev$. The shaded region on the right is excluded by the chargino
search, while the shaded region on the left is excluded if a neutral
LSP is required.}
\label{fig:m0minit}
\end{figure}

We may also consider variations in the high scale $Q_0$ where the
supersymmetry breaking parameters are assumed to be generated.  So far
we have assumed $Q_0 = \mgut$.  It is interesting to consider the
effects of assuming that the boundary conditions are specified at a
different scale, e.g., the string or Planck scale. To investigate
this, we have taken a simple approach, and evolved the gauge couplings
up to some fixed $Q_0$, set the supersymmetry breaking parameters at
that scale, and then evolved them down to the weak scale.  The minimal
field content is assumed throughout the RG evolution; in particular,
no additional GUT particle content is assumed for $Q_0 > \mgut$, and
the unification of gauge couplings at $\mgut$ is unexplained.  In
Fig.~\ref{fig:m0minit} we show contours of constant $c$ in the
$(m_0,Q_0)$ plane. Just as in Fig.~\ref{fig:m0M1/2}, the fine-tuning
parameter $c$ is determined successively, for increasing $m_0$, by
$c_{\mu_0}$, $c_{M_{1/2}}$ and $c_{m_0}$.  We see that increasing the
scale $Q_0$ also allows even larger scalar masses. For example the
requirements $c\leq 50$ and $Q_0<\mplanck$ allow $m_0$ as large as 2.9
TeV.

\begin{figure}[tbp]
\postscript{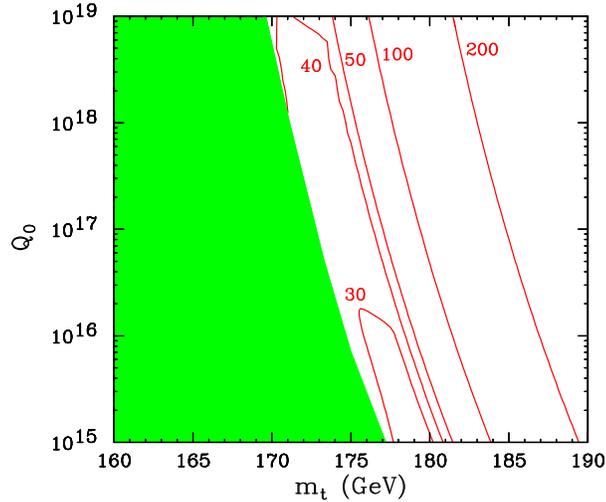}{0.49}
\caption
{Contours of constant fine-tuning $c$ in the $( \mt, \minit)$ plane
for $m_0=2 \tev$, $\mgaugino = 300 \gev$, $A_0=0$, $\tb = 10$, and
$\mu>0$. The shaded region is excluded from the chargino mass limit.}
\label{fig:minitmt}
\end{figure}

In Fig.~\ref{fig:minitmt} we show contours of constant $c$ in the
$(m_t,Q_0)$ plane. As expected, smaller values of $m_t$ can be
compensated for by larger evolution intervals, and vice versa. Notice,
however, that varying $m_t$ within its current experimental
uncertainty leads to changes in $c$ that are as large as those caused
by varying $Q_0$ by several orders of magnitude.

\section{Implications for Supersymmetry Searches}
\label{sec:implications}

We have seen that the naturalness bound on $m_0$ (i.e., the typical
sfermion mass) may be as large as a few TeV.  In this section, we
discuss the implication of these results for the superpartner spectrum
and, in particular, the discovery prospects for scalar superpartners
at future colliders.  (Of course, it is clear that such heavy scalars
also drastically reduce the size of supersymmetric effects in low
energy experiments, but we will not address this further.)

The implications for sleptons are fairly straightforward.  Sleptons
have small Yukawa couplings (with the possible exception of staus for
large $\tb$), and so their masses are virtually RG-invariant, with
$m_{\tilde{l}} \approx m_0$ in this scenario.  Multi-TeV sleptons are
beyond the kinematic limit $m_{\tilde{l}} < \sqrt{s}/2$ of all
proposed linear colliders.  They will also escape detection at hadron
colliders, as they are not strongly produced, and will not be produced
in large numbers in the cascade decays of strongly interacting
superparticles.

We now turn to squarks.  Multi-TeV squarks will, of course, also evade
proposed linear colliders.  Traditionally, however, it has been
expected that future hadron colliders, particularly the LHC, will
discover all squarks in the natural region of minimal supergravity
parameter space~\cite{Snowmass}.  This conclusion is based on the
expectation that all squarks have masses $\alt 1-2~{\rm TeV}$. In
Fig.~\ref{fig:msq}, we present contours for
$m_{\tilde{u}_L}$.\footnote{One-loop corrections~\cite{PBMZ} are
included in all superpartner masses.}  (All first and second
generation squarks are nearly degenerate.) In the same figure, we have
also included contours of the fine-tuning parameter $c$.  For $c \leq
50$, we see that squark masses of $m_{\tilde{u}_L} = 2.2~{\rm TeV}$
are allowed, and more generally, the parameter space with multi-TeV
squarks is as natural as parameter space with $m_{\tilde{u}_L}\alt
1~{\rm TeV}$.  Recall also that these mass limits may be extended to
as large as $\sim 3 \tev$ for variations of $\mt$ within its current
bounds.  Squarks of mass $\sim 2 \tev$ may be detected at the LHC with
large integrated luminosities of several $100 \ifb$.  However, squarks
with masses significantly beyond 2 TeV are likely to escape detection
altogether~\cite{Hinchliffe}.

\begin{figure}[tbp]
\postscript{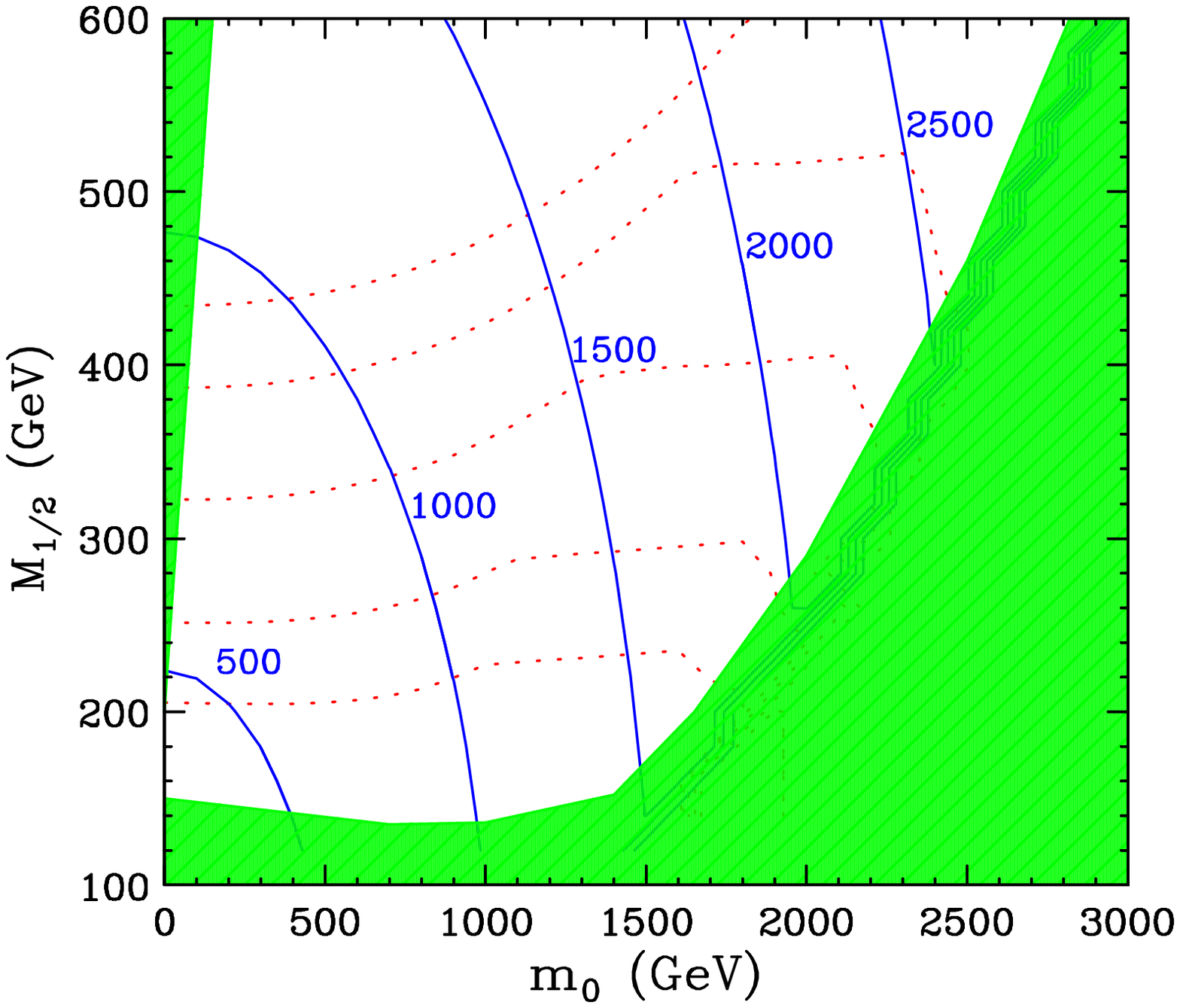}{0.49}
\caption
{Contours of constant squark mass $m_{\tilde{u}_L}$ (solid) in the
$(m_0, \mgaugino)$ plane for $A_0=0$, $\tb = 10$, $\mu>0$, and $\mt =
174$ GeV. Fine-tuning contours (dotted) are also presented for $c =
20$, 30, 50, 75, and 100 (from below). The bottom and right shaded
region is excluded by the chargino mass limit of 90 GeV.  The top left
region is also excluded if a neutral LSP is required.}
\label{fig:msq}
\end{figure}

Since the top squarks and left-handed bottom squark interact strongly
through $y_t$, they are lighter than the other squarks.  For small
$\tan\beta$, and small $\mgaugino$ and $A_0$ parameters, their masses
at the focus point of $m_{H_u}^2$ are given by
\begin{eqnarray}
m_{U_3}^2 \simeq \frac{1}{3}m_0^2, \quad
m_{Q_3}^2 \simeq \frac{2}{3}m_0^2 \ .
\end{eqnarray}
In general, these squark masses, particularly the stop masses, may
also be influenced by left-right mixing.  However, because naturalness
constrains the $A$ and $\mu$ parameters to be of order the weak
scale, left-right mixing effects are sub-leading for multi-TeV $m_0$.
As a result, the lighter (heavier) stop is mostly right-handed
(left-handed), and the lighter (heavier) sbottom is mostly left-handed
(right-handed).  For large $\tan\beta$, $y_b$ also suppresses third
generation squark masses.  For example, for $y_b=y_t$, we obtain
\begin{eqnarray}
m_{U_3}^2 \simeq m_{D_3}^2 \simeq m_{Q_3}^2 
\simeq \frac{1}{3}m_0^2 \ .
\end{eqnarray}

In Figs.~\ref{fig:mstop} and \ref{fig:msbot} we present the masses of
the stops and sbottoms, respectively.  By comparing with
Fig.~\ref{fig:msq}, we see that the stops are always lighter than the
first and second generation squarks.  The lighter sbottom and heavier
stop are nearly degenerate, since they are (approximately) in the same
SU(2)$_L$ doublet.  The heavier sbottom, which is mostly the
right-handed sbottom, may also become significantly lighter than the
first two generation squarks for large $\tan\beta$.  Therefore, in the
multi-TeV $m_0$ scenario, stop and sbottom production will be the most
promising modes for squark discovery at the LHC.

\begin{figure}[tbp]
\postscript{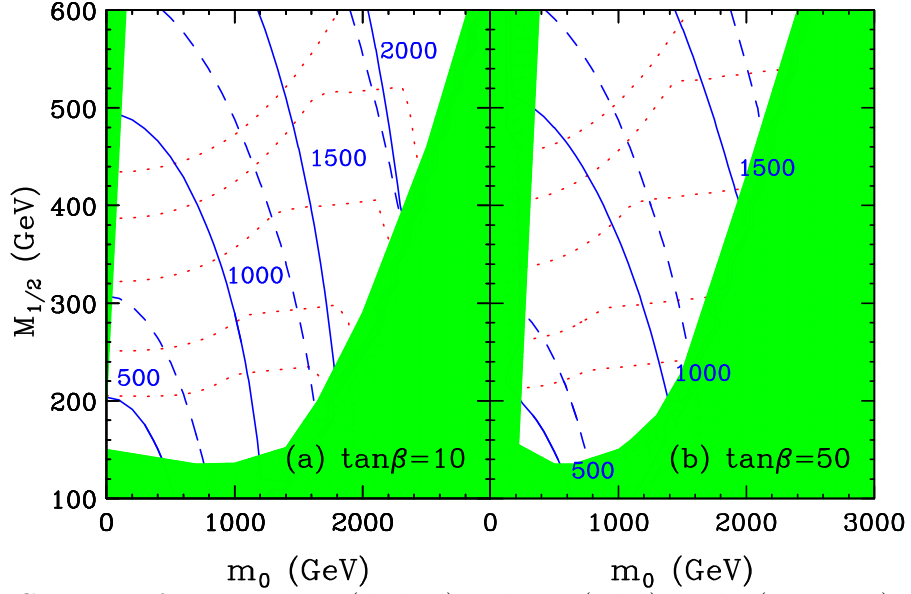}{0.74}
\caption
{Contours of constant $m_{\tilde{t}_1}$ (dashed) and $m_{\tilde{t}_2}$
(solid) in the $(m_0, \mgaugino)$ plane for (a) $\tb=10$ and (b)
$\tb=50$, $A_0=0$, $\mu>0$, and $\mt = 174$ GeV.  Fine-tuning contours
(dotted) are also presented for $c = 20$, 30, 50, 75, and 100 (from
below). The bottom and right shaded region is excluded by the chargino
mass limit of 90 GeV.  The top left region is also excluded if a
neutral LSP is required.}
\label{fig:mstop}
\end{figure}

\begin{figure}[tbp]
\postscript{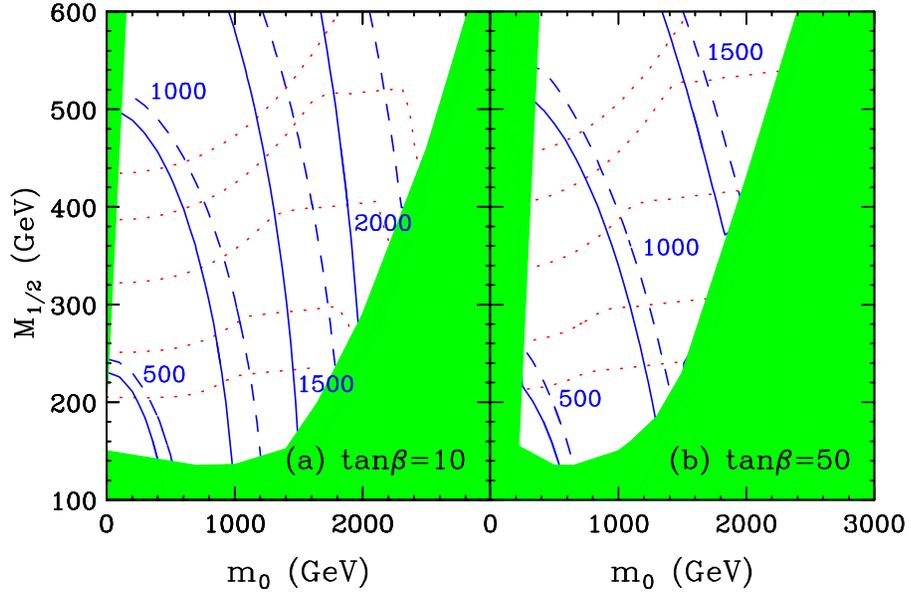}{0.74}
\caption
{As in Fig.~\ref{fig:mstop}, but for $m_{\tilde{b}_1}$ (dashed) and
$m_{\tilde{b}_2}$ (solid). }
\label{fig:msbot}
\end{figure}

In contrast to the sfermions, gauginos cannot be very heavy in this
scenario. This fact is explicit in Fig.~\ref{fig:c_mgaugino}: for
large $M_{1/2}$, the fine-tuning coefficient $c_{M_{1/2}}$ becomes
unacceptably large, irrespective of $\tan\beta$.  As a result,
naturalness requires fairly light gauginos.  For example, the
constraint $c\leq 50$ implies $M_{1/2}\lesssim 400~{\rm GeV}$,
corresponding to $M_1\lesssim 160~{\rm GeV}$, $M_2\lesssim 320~{\rm
GeV}$, and $M_3\lesssim 1.2~{\rm TeV}$.  Such gauginos will be
produced in large numbers at the LHC, and will be discovered in
typical scenarios.\footnote{In some scenarios, however, the detection
of all gauginos may be challenging. This is particularly true in
scenarios with degeneracies, such as the Wino LSP
scenario~\cite{WinoLSPexp}. To our knowledge, the detectability of all
gauginos (not including the invisible LSP) in such scenarios at the
LHC remains an open question.}

It is also interesting to consider the implications of the focus point
for Higgs masses.  In supersymmetric models with minimal field
content, the lightest Higgs boson mass $m_h$ is bounded by $m_Z$ at
tree level.  However, this upper bound may be significantly violated
by radiative corrections~\cite{RCtoHiggs}.  In particular, top-stop
loop contributions, approximately proportional to
$\ln(m_{\tilde{t}}/m_t)$, may be important.  Since the focus point
allows heavy stops, one may wonder if this affects the upper bound on
the lightest Higgs mass.  To answer this question, we have calculated
the one-loop radiative corrections to the lightest Higgs mass. The
result is shown in Fig.~\ref{fig:mh}; we emphasize the $A_0$
dependence, as the radiative correction is sensitive to left-right
mixing through the $A$ parameter.  For the region with small
fine-tuning parameter, say, $c\leq 50$, we find $m_h < 118 \gev$.  It
is important that a large $A$ parameter is forbidden by naturalness,
as this suppresses left-right stop mixing, which usually significantly
increases $m_h$.  Therefore, even in the focus point scenario with
multi-TeV squarks, Run II of the Tevatron will probe much of parameter
space in its search for the lightest Higgs boson \cite{Higgs report}.

\begin{figure}[tbp]
\postscript{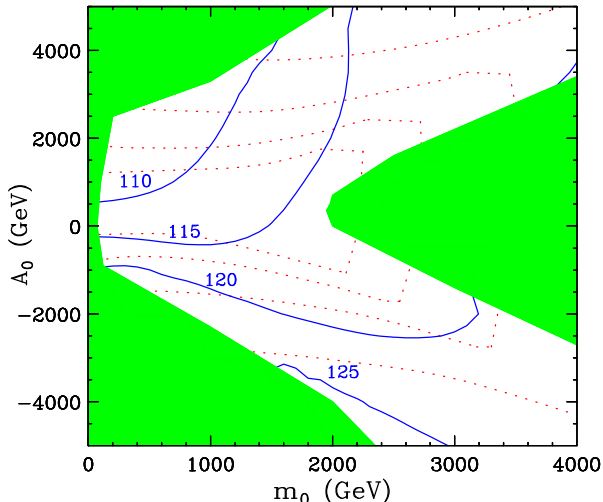}{0.49}
\caption
{Contours of constant $m_h$ (solid) in the $(m_0, A_0)$ plane for
$\mgaugino=300$ GeV, $\tb = 10$, $\mu>0$, and $\mt = 174$
GeV. Fine-tuning contours (dotted) are also presented for $c = 50$,
100, 200, and 500 (see Fig.~\ref{fig:m0A0}a).}
\label{fig:mh}
\end{figure}

Discovery of the heavy Higgs scalars is more challenging.  At
tree-level, the masses of the heavy Higgs scalars are approximately
given by

\begin{eqnarray}
m_A \simeq m_H \simeq m_{H^\pm} \simeq
\sqrt{m_{H_u}^2+m_{H_d}^2-2\mu^2} \ .
\end{eqnarray}
Although $m_{H_u}^2$ and $\mu^2$ are always bounded by naturalness,
for moderate $\tan\beta$, $m_{H_d}^2$ is only weakly bounded by
naturalness.  For negligible $y_b$, $m_{H_d}^2$ does not participate
in the focus point behavior and is roughly RG-invariant. For larger
$\tan\beta$, on the other hand, $m_{H_d}^2$ may be significantly
suppressed by $y_b$ during RG evolution.  In particular, in the case
of $\tan\beta\simeq m_t/m_b$, there is an approximate symmetry under
interchanging $H_u \leftrightarrow H_d$ and $U_3 \leftrightarrow D_3$,
and so $m_{H_d}^2$ also has a focus point near the weak scale.  In
Fig.~\ref{fig:mA}, we present the pseudoscalar Higgs mass $m_A$ in the
$(m_0, M_{1/2})$ plane.  For $\tan\beta =10$, $m_{H_d}^2\simeq m_0^2$,
$m_A \sim m_0$, and detection of the heavy Higgses at the LHC or
proposed linear colliders becomes impossible for large $m_0$.
However, for large $\tan\beta$, $m_{H_d}^2$ is suppressed by $y_b$,
and $m_A \sim {\cal O} (100 \gev)$.  For large $\tan\beta$, heavy
Higgses with masses of several hundred GeV may be found at the LHC
through the decays $H,A \to \tau\bar{\tau}$~\cite{Higgstau}.

\begin{figure}[tbp]
\postscript{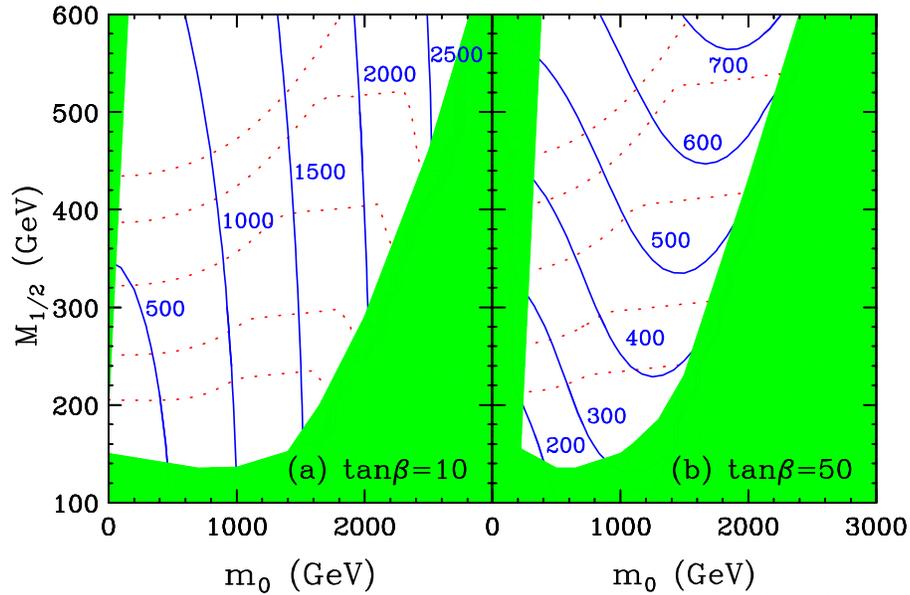}{0.74}
\caption
{Contours of constant heavy Higgs mass $m_A \approx m_H \approx
m_{H^{\pm}}$ (solid) in the $(m_0, \mgaugino)$ plane for (a) $\tb=10$
and (b) $\tb=50$, $A_0=0$, $\mu>0$, and $\mt = 174$ GeV. Fine-tuning
contours (dotted) are also presented for $c = 20$, 30, 50, 75, and 100
(from below).}
\label{fig:mA}
\end{figure}

\section{Summary and Discussion}
\label{sec:summary}

In this paper, we have explored the existence of focus points in the
RG behavior of supersymmetry breaking parameters and their
implications for naturalness and experimental searches for
supersymmetry.

For the experimentally measured top quark mass, the supersymmetry
breaking up-type Higgs mass parameter $m_{H_u}^2$ has a focus point at
the scale $Q\sim {\cal O}(100~{\rm GeV})$ in a class of models which
includes minimal supergravity. The value of $m_{H_u}^2$ at the
weak scale is therefore highly insensitive to the universal
scalar mass $m_0$ at the GUT scale.  We have also seen that this focus
point behavior exists for all values of $\tb \agt 5$.

Since $m_{H_u}^2$ plays an important role in the determination of the
weak scale, this focus point behavior affects the naturalness of
electroweak symmetry breaking in minimal supergravity.  In particular,
because a large $m_0$ can result in a reasonably small $m_{H_u}^2$,
naturalness constraints on $m_0$ are not as severe as typically
expected.  To discuss this issue quantitatively, we have calculated
the fine-tuning parameter $c$, which is determined by the sensitivity
of the weak scale to fractional variations of the fundamental
parameters.  As we have seen, in regions of parameter space with
$m_0\sim {\rm 2-3~TeV}$ this fine-tuning parameter may be as small as
in regions with $m_0\lesssim {\rm 1~TeV}$.  As a result, multi-TeV
sfermions are as natural as sfermions lighter than ${\rm 1~TeV}$.  We
note that the region of multi-TeV scalars and light gauginos {\em and}
Higgsinos is also somewhat preferred by gauge coupling unification in
minimal SU(5) \cite{BMP,CPP,BMPZ-GM}, as well as $b$-$\tau$ Yukawa
unification at moderate to large $\tan\beta$ \cite{MP,Raby}.  The
discovery of squarks and sleptons at the LHC and proposed linear
colliders may therefore be extremely challenging, and may require some
even more energetic machines, such as muon or very large hadron
colliders.

In our analysis, we did not include the Yukawa couplings, notably
$y_t$, and gauge couplings in the calculation of the the fine-tuning
parameter $c$.  In Fig.~\ref{fig:c_yt}, we present the sensitivity
coefficient $c_{y_t}$ for the (GUT scale) top Yukawa coupling.  For
$m_0\geq 1~{\rm TeV}$, $c_{y_t}$ is always larger than 70, and so if
$c_{y_t}$ is included in the calculation of the fine-tuning parameter
$c$, the naturalness bound on $m_0$ becomes much more stringent and
$m_0\gtrsim 1~{\rm TeV}$ is disfavored.  The inclusion of $y_t$ and
other standard model parameters in the fine-tuning calculation would
thus lead to significantly different conclusions.
\begin{figure}[tbp]
\postscript{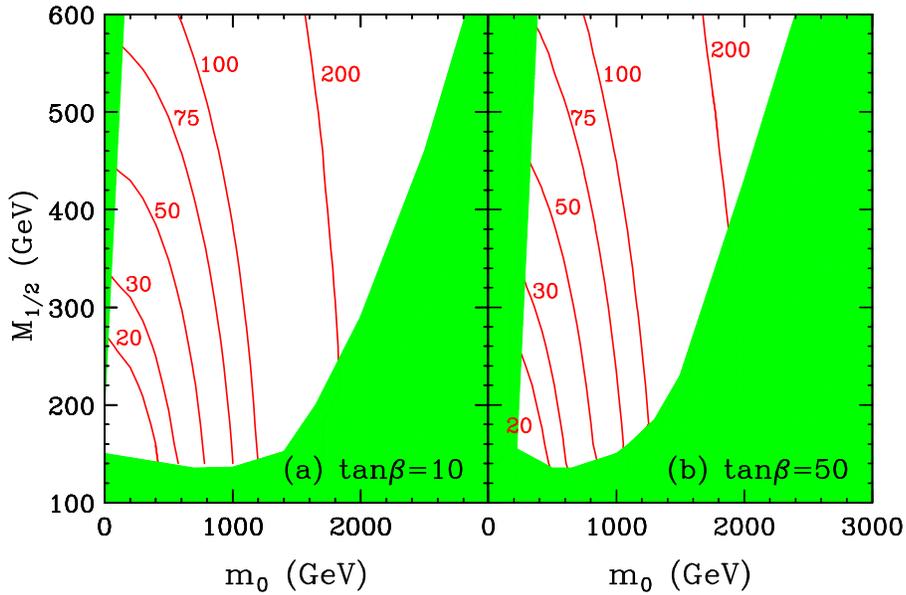}{0.74}
\caption
{As in Fig.~\ref{fig:c_m0}, but for the sensitivity parameter
$c_{y_t}$. }
\label{fig:c_yt}
\end{figure}

A definitive resolution to this question of whether or not to include
variations of standard model parameters in fine-tuning calculations
cannot, we believe, be achieved without a more complete understanding
of the fundamental theories of flavor and supersymmetry breaking.
Without such knowledge, any discussion necessarily becomes somewhat
philosophical.  Nevertheless, several remarks are in order.

First, it is sometimes argued that one should not consider variations
with respect to parameters that have been measured or are highly
correlated with measured quantities.  According to this view, standard
model parameters such as the strong coupling constant, and possibly
also $y_t$, should not be included among the $a_i$. We do not
subscribe to this view.\footnote{Note that the exclusion of standard
model parameters from the fundamental parameters $a_i$ does not imply
that current experimental data are ignored in the calculation of
fine-tuning.  All experimental data are used in step ({\it ii}) of the
fine-tuning prescription to specify the physical hypersurface of
parameter space.}  If in the future the Higgsino mass $\mu$ is
measured to be 10 TeV, given our current notions of naturalness, we
believe this should be considered fine-tuned, irrespective of the
accuracy with which the Higgsino mass is measured. Of course, if this
were the case, the fact that a 10 TeV $\mu$ parameter is realized in
nature would be a strong motivation to consider alternative, and
perhaps more fundamental, theoretical frameworks in which a 10 TeV
$\mu$ parameter is not unnatural.

There are, however, other considerations which favor the exclusion of
standard model parameters from the list of $a_i$.  As noted in
Sec.~\ref{sec:natural}, we are interested in the naturalness of the
supersymmetric explanation of the gauge hierarchy.  We should not
require that supersymmetry also solve the problem of flavor.  In fact,
in many supergravity frameworks, the supersymmetry breaking parameters
and the Yukawa couplings are expected to be determined independently.
For example, in hidden sector scenarios, the supersymmetry breaking
parameters are determined in one sector, while the Yukawa couplings
are fixed in some other sector and by a completely independent
mechanism. In this case, it seems reasonable to assume that $y_t$ is
fixed to its observed value in some sector not connected to
supersymmetry breaking, and we therefore should not consider
variations with respect to it.

Finally, it is worth noting that there are several possible scenarios
in which it is clear that $c_{y_t}$ should not be included in $c$ or
is at least negligibly small.  One possibility is that $y_t$ may
evolve from some higher scale, such as the Planck scale, to a fixed or
focus point at the GUT scale.  The weak scale is then highly
insensitive to variations in the truly fundamental parameter, i.e.,
$y_t$ at the Planck scale.  Alternatively, the top Yukawa coupling may
arise as a renormalizable operator with coefficient determined by a
correlation function of string vertex operators. The coupling $y_t$
would then be fixed to its current value (or possibly one of a
discrete set of values), and it is again inappropriate to consider
continuous variations with respect to $y_t$.  Note that in both of
these scenarios, $y_t$ may receive additional contributions from
non-renormalizable operators of the form $\delta y_t \sim g \epsilon$,
where $g$ is a coupling constant, and $\epsilon$ is some small
expansion parameter, such as $v/M_{Pl}$, where $v$ is some vacuum
expectation value.  In this case, $c_g$ and $c_{\epsilon}$ should be
included in the definition of fine-tuning, but they will be negligible
for small $\epsilon$.

\section*{Acknowledgments}

We thank J.~Bagger, G.~Giudice, and C.~Wagner for discussions.  This
work was supported in part by the Department of Energy under contracts
DE--FG02--90ER40542 and DE--AC02--76CH03000, by the National Science
Foundation under grant PHY--9513835, through the generosity of Frank
and Peggy Taplin (JLF), and by a Marvin L.~Goldberger Membership (TM).

\appendix

\section*{Determination of Focus Point Scale for $\bold{\lowercase{y_b
\ll y_t}}$ and $\bold{\lowercase{y_b = y_t}}$}
\label{app:largetb}

In this appendix, we discuss the focus point analytically at one-loop
for the two cases $y_b \ll y_t$ and $y_b = y_t$.  Solutions to the RG
equations are well-known for these two cases~\cite{RGEanalysis}.  For
both cases, we derive a closed form expression involving the gauge
couplings and $\mt$ that must be satisfied if the focus point scale
$Q_{\rm F}$ is to be at the weak scale.  We also show that if the
focus point is at the weak scale for $y_b\ll y_t$, it is also at the
weak scale for $y_b=y_t$.

In the analysis of the $y_b=y_t$ case, we neglect the effects of
$y_\tau$ and the hypercharge differences in the $y_t$ and $y_b$ RG
equations, so $y_t$ and $y_b$ remain degenerate throughout their RG
evolution.  The validity of these approximations is verified only by
the numerical results in Sec.~\ref{sec:focus}. In addition, the
intermediate case, where $y_b \ne y_t$ but $y_b$ may not be neglected,
is not considered.  However, the following analysis may be helpful in
understanding the numerical results, and in particular, the behavior
of Figs.~\ref{fig:run} and \ref{fig:focus}.

Let us define
\begin{eqnarray}
\tilde{\alpha}_a &\equiv& \frac{g_a^2}{16\pi^2}, \ \ a=1,2,3,
\\
\tilde{\alpha}_y &\equiv& \frac{y_t^2}{16\pi^2} \ .
\end{eqnarray}
These quantities obey the following RG equations:
\begin{eqnarray}
\frac{d \tilde{\alpha}_a}{d \ln Q} &=&
2 b_a \tilde{\alpha}_a^2 \ ,
\\
\frac{d \tilde{\alpha}_y}{d \ln Q} &=&
2 \biggl( s \tilde{\alpha}_y - \sum_a r_a \tilde{\alpha}_a  \biggr)
\tilde{\alpha}_y \ .
\end{eqnarray}
In the minimal supersymmetric standard model,
$(b_1,b_2,b_3)=(11,1,-3)$, $(r_1,r_2,r_3)=(13/9,3,16/3)$, and $s=6$
and 7 for $y_b \ll y_t$ and $y_b=y_t$, respectively.  The solution for
$\tilde{\alpha}_y$ is
\begin{eqnarray}
\tilde{\alpha}_y (Q) =
\frac{\tilde{\alpha}_y(Q_0)E(Q)}{1-2s\tilde{\alpha}_y(Q_0)F(Q)} \ ,
\end{eqnarray}
where
\begin{eqnarray}
E(Q) &=&
\prod_a
\left[ 1 - 2 b_a \tilde{\alpha}_a(Q_0) \ln(Q/Q_0) \right]^{r_a/b_a} 
\ ,
\\
F(Q) &=&
\int_{\ln Q_0}^{\ln Q} E(Q') d\ln Q' \ .
\end{eqnarray}
We find then that
\begin{eqnarray}
e^{sI(Q)} \equiv
\exp \left( 2s \int_{\ln Q_0}^{\ln Q} \tilde{\alpha}_y(Q') d\ln Q'
\right)
= \frac{1}{1-2s\tilde{\alpha}_y(Q_0)F(Q)}
= 1 + \frac{2s\tilde{\alpha}_y(Q)F(Q)}{E(Q)} \ .
\end{eqnarray}

For a universal scalar mass, the conditions for the focus point (see
Sec.~\ref{sec:focus}) are

\begin{eqnarray}
e^{sI} =
\left\{ \begin{array}{l}
1/3 \ , \ {\rm for}~y_b\ll y_t
\\
2/9 \ , \ {\rm for}~y_b=y_t
\end{array} \right.  \ .
\end{eqnarray}
We see that these are simultaneously satisfied at $Q=m_t$ if
\begin{eqnarray}
\frac{\tilde{\alpha}_y(m_t)F(m_t)}{E(m_t)} = - \frac{1}{18} \ .
\end{eqnarray}
In terms of the top quark mass, this requirement corresponds to
\begin{eqnarray}
\frac{1}{16\pi^2} \left[\frac{m_t(m_t)}{v \sin\beta} \right]^2
\frac{F(m_t)}{E(m_t)} = - \frac{1}{18} \ ,
\label{closedform}
\end{eqnarray}
where $m_t(m_t)$ is the running top quark mass, and $v = 174~{\rm
GeV}$. 

Given fixed gauge coupling constants, Eq.~(\ref{closedform}) specifies
which top quark mass will place the focus point at the weak
scale.  For $Q_0=M_{\rm GUT}=2\times 10^{16}~{\rm GeV}$ and $\alpha
(M_{\rm GUT})=1/24$, we obtain $F\simeq -130$ and $E\simeq 13$ at
$Q=174~{\rm GeV}$.  For $\sin\beta\simeq 1$, the requirement is then
$m_t(m_t)\simeq 160~{\rm GeV}$, which is very close to the running
mass corresponding to the physical pole mass $m_t\approx 174~{\rm
GeV}$. Therefore, for the experimentally measured top quark mass, the
focus point of $m_{H_u}^2$ is close to the weak scale for
$y_b=y_t$ and also for $y_b\ll y_t$.

\end{document}